\documentclass[ALICE,manyauthors]{cernphprep}
\usepackage[comma,square,numbers,sort&compress]{natbib}
\usepackage[T1]{fontenc}
\usepackage{orcidlink}
\usepackage{hyperref}
\usepackage{booktabs}
\usepackage{lineno}
\usepackage{color}
\usepackage{tabularray}
\usepackage{multirow}
\usepackage{cleveref} 
\crefname{equation}{Eq}{Eqs.}
\crefrangelabelformat{equation}{(#3#1#4--#5#2#6)}
\newcolumntype{Y}{>{\centering\arraybackslash}X}

\RequirePackage[normalem]{ulem} 
\RequirePackage{color}\definecolor{RED}{rgb}{1,0,0}\definecolor{BLUE}{rgb}{0,0,1} 

\begin{document}
%

\newcommand{\red}{\textcolor{red}}
\newcommand{\blue}{\textcolor{blue}}


\newcommand{\pp}           {pp\xspace}
\newcommand{\ppbar}        {\mbox{$\mathrm {p\overline{p}}$}\xspace}
\newcommand{\XeXe}         {\mbox{Xe--Xe}\xspace}
\newcommand{\PbPb}         {\mbox{Pb--Pb}\xspace}
\newcommand{\pA}           {\mbox{pA}\xspace}
\newcommand{\pPb}          {\mbox{p--Pb}\xspace}
\newcommand{\AuAu}         {\mbox{Au--Au}\xspace}
\newcommand{\dAu}          {\mbox{d--Au}\xspace}

\newcommand{\s}            {\ensuremath{\sqrt{s}}\xspace}
\newcommand{\snn}          {\ensuremath{\sqrt{s_{\mathrm{NN}}}}\xspace}
\newcommand{\pt}           {\ensuremath{p_{\rm T}}\xspace}
\newcommand{\mt}           {\ensuremath{m_{\mathrm{T}}}\xspace}
\newcommand{\meanptref}    {\ensuremath{\langle p_{\rm T}\rangle^{0-5\%}}\xspace}
\newcommand{\meanpt}       {\ensuremath{\langle p_{\rm T}\rangle}\xspace}
\newcommand{\normmeanpt}   {\ensuremath{\langle p_{\rm T}\rangle/\langle p_{\rm T}\rangle^{0-5\%}}\xspace}
\newcommand{\ycms}         {\ensuremath{y_{\rm CMS}}\xspace}
\newcommand{\ylab}         {\ensuremath{y_{\rm lab}}\xspace}
\newcommand{\etarange}[1]  {\mbox{$\left | \eta \right |~<~#1$}}
\newcommand{\yrange}[1]    {\mbox{$\left | y \right |~<~#1$}}
\newcommand{\dndy}         {\ensuremath{\mathrm{d}N/\mathrm{d}y}\xspace}
\newcommand{\dndeta}       {\ensuremath{\mathrm{d}N_\mathrm{ch}/\mathrm{d}\eta}\xspace}
\newcommand{\avdndeta}     {\ensuremath{\langle\dndeta\rangle}\xspace}
\newcommand{\dNdy}         {\ensuremath{\mathrm{d}N_\mathrm{ch}/\mathrm{d}y}\xspace}
\newcommand{\Npart}        {\ensuremath{N_\mathrm{part}}\xspace}
\newcommand{\avNpart}      {\ensuremath{\langle N_\mathrm{part}\rangle}\xspace}
\newcommand{\Ncoll}        {\ensuremath{N_\mathrm{coll}}\xspace}
\newcommand{\dEdx}         {\ensuremath{\textrm{d}E/\textrm{d}x}\xspace}
\newcommand{\avEZDC}       {\ensuremath{\langle E_{\mathrm{ZDC}} \rangle}\xspace}
\newcommand{\avEZN}        {\ensuremath{\langle E_{\mathrm{N}} \rangle}\xspace}
\newcommand{\avEZP}        {\ensuremath{\langle E_{\mathrm{P}} \rangle}\xspace}
\newcommand{\normdndeta}   {\ensuremath{\langle\dndeta\rangle/\langle\dndeta\rangle^{0-5\%}}\xspace}
\newcommand{\cs}           {\ensuremath{c_{\mathrm{s}}}\xspace}
\newcommand{\scs}          {\ensuremath{c_{\mathrm{s}}^{2}}\xspace}
\newcommand{\ntracklets}   {\ensuremath{N_{\mathrm{tracklets}}}\xspace}
\newcommand{\nch}          {\ensuremath{N_{\mathrm{ch}}}\xspace}
\newcommand{\et}           {\ensuremath{E_{\mathrm{T}}}\xspace}
\newcommand{\nchnorm}      {\ensuremath{\langle N_{\mathrm{ch}} \rangle^{\mathrm{norm}}}\xspace}
\newcommand{\nchknee}      {\ensuremath{\langle N_{\mathrm{ch}} \rangle^{\mathrm{knee}}}\xspace}
\newcommand{\ptnorm}       {\ensuremath{\langle p_{\mathrm{T}}\rangle^{\mathrm{norm}}}\xspace}
\newcommand{\dndetanorm}   {\ensuremath{\avdndeta^{\mathrm{norm}}}\xspace}
\newcommand{\dndetanormknee}   {\ensuremath{\avdndeta^{\mathrm{norm}}_{\mathrm{knee}}}\xspace}
\newcommand{\sigmanormknee}    {\ensuremath{\sigma^{\mathrm{norm}}_{\mathrm{knee}}}\xspace}

\newcommand{\mmpt}{\ensuremath{\langle[\pt]\rangle}\xspace}
\newcommand{\mptk}[1]{\ensuremath{[\pt^{(#1)}]}\xspace}
\newcommand{\mmptk}[1]{\ensuremath{\langle[\pt^{(#1)}]\rangle}\xspace}
\newcommand{\stdskew}        {\ensuremath{\gamma_{\langle [p_{\mathrm{T}}]\rangle}}\xspace}
\newcommand{\intskew}        {\ensuremath{\Gamma_{\langle [p_{\mathrm{T}}]\rangle}}\xspace}
\newcommand{\kur}        {\ensuremath{\kappa_{\langle [p_{\mathrm{T}}]\rangle}}\xspace}

\newcommand{\ktwonorm}        {\ensuremath{k_2/k_2^{0-5\%}\xspace}}
\newcommand{\kthreenorm}      {\ensuremath{k_3/k_3^{0-5\%}\xspace}}
\newcommand{\stdskewnorm}     {\ensuremath{\stdskew/\stdskew^{0-5\%}\xspace}}
\newcommand{\intskewnorm}     {\ensuremath{\intskew/\intskew^{0-5\%}\xspace}}
\newcommand{\kurnorm}         {\ensuremath{\kur/\kur^{0-5\%}\xspace}}

\newcommand{\ktwnorm}        {\ensuremath{k_2^{\mathrm{norm}}\,\,}}
\newcommand{\kthnorm}        {\ensuremath{k_3^{\mathrm{norm}}\,\,}}
\newcommand{\stdsknorm}      {\ensuremath{\stdskew^{\mathrm{norm}}\xspace}}
\newcommand{\intsknorm}      {\ensuremath{\intskew^{\mathrm{norm}}\xspace}}
\newcommand{\kunorm}         {\ensuremath{\kur^{\mathrm{norm}}\xspace}}

\newcommand{\nineH}        {$\sqrt{s}~=~0.9$~Te\kern-.1emV\xspace}
\newcommand{\seven}        {$\sqrt{s}~=~7$~Te\kern-.1emV\xspace}
\newcommand{\twoH}         {$\sqrt{s}~=~0.2$~Te\kern-.1emV\xspace}
\newcommand{\twosevensix}  {$\sqrt{s}~=~2.76$~Te\kern-.1emV\xspace}
\newcommand{\five}         {$\sqrt{s}~=~5.02$~Te\kern-.1emV\xspace}
\newcommand{\thirteen}    {$\sqrt{s}~=~13$~Te\kern-.1emV\xspace}
\newcommand{\twosevensixnn}{$\sqrt{s_{\mathrm{NN}}}~=~2.76$~Te\kern-.1emV\xspace}
\newcommand{\fivenn}       {$\sqrt{s_{\mathrm{NN}}}~=~5.02$~Te\kern-.1emV\xspace}
\newcommand{\LT}           {L{\'e}vy-Tsallis\xspace}
\newcommand{\GeVc}         {Ge\kern-.1emV/$c$\xspace}
\newcommand{\MeVc}         {Me\kern-.1emV/$c$\xspace}
\newcommand{\TeV}          {Te\kern-.1emV\xspace}
\newcommand{\GeV}          {Ge\kern-.1emV\xspace}
\newcommand{\MeV}          {Me\kern-.1emV\xspace}
\newcommand{\GeVmass}      {Ge\kern-.2emV/$c^2$\xspace}
\newcommand{\MeVmass}      {Me\kern-.2emV/$c^2$\xspace}
\newcommand{\lumi}         {\ensuremath{\mathcal{L}}\xspace}
\newcommand{\gevc}[1]      {\ensuremath{#1 \text{\,GeV/$c$}}\xspace}
\newcommand{\mevc}[1]      {\ensuremath{#1 \text{\,MeV/$c$}}\xspace}
\newcommand{\sppt}[1]      {\ensuremath{\sqrt{s} = #1 \text{\,TeV}}\xspace}

\newcommand{\ITS}          {\rm{ITS}\xspace}
\newcommand{\TOF}          {\rm{TOF}\xspace}
\newcommand{\ZDC}          {\rm{ZDC}\xspace}
\newcommand{\ZDCs}         {\rm{ZDCs}\xspace}
\newcommand{\ZNA}          {\rm{ZNA}\xspace}
\newcommand{\ZNC}          {\rm{ZNC}\xspace}
\newcommand{\SPD}          {\rm{SPD}\xspace}
\newcommand{\SDD}          {\rm{SDD}\xspace}
\newcommand{\SSD}          {\rm{SSD}\xspace}
\newcommand{\TPC}          {\rm{TPC}\xspace}
\newcommand{\TRD}          {\rm{TRD}\xspace}
\newcommand{\VZERO}        {\rm{V0}\xspace}
\newcommand{\VZEROA}       {\rm{V0A}\xspace}
\newcommand{\VZEROC}       {\rm{V0C}\xspace}
\newcommand{\Vdecay} 	   {\ensuremath{V^{0}}\xspace}

\newcommand{\hadrons}      {\ensuremath{\mathrm{h}^{\pm}}\xspace}
\newcommand{\pion}         {\ensuremath{\pi}\xspace}
\newcommand{\kaon}         {\ensuremath{\textrm{K}}\xspace}
\newcommand{\pr}           {\ensuremath{\textrm{p}}\xspace}
\newcommand{\prx}          {\ensuremath{\mathrm{p}(\overline{\mathrm{p}}})\xspace}
\newcommand{\ee}           {\ensuremath{e^{+}e^{-}}} 
\newcommand{\pip}          {\ensuremath{\pi^{+}}\xspace}
\newcommand{\pim}          {\ensuremath{\pi^{-}}\xspace}
\newcommand{\kapm}         {\ensuremath{\mathrm{K}^{\pm}}}
\newcommand{\kap}          {\ensuremath{\rm{K}^{+}}\xspace}
\newcommand{\kam}          {\ensuremath{\rm{K}^{-}}\xspace}
\newcommand{\pbar}         {\ensuremath{\rm\overline{p}}\xspace}
\newcommand{\kzero}        {\ensuremath{{\rm K}^{0}_{\rm{S}}}\xspace}
\newcommand{\lmb}          {\ensuremath{\Lambda}\xspace}
\newcommand{\almb}         {\ensuremath{\overline{\Lambda}}\xspace}
\newcommand{\Om}           {\ensuremath{\Omega^-}\xspace}
\newcommand{\Mo}           {\ensuremath{\overline{\Omega}^+}\xspace}
\newcommand{\X}            {\ensuremath{\Xi^-}\xspace}
\newcommand{\Ix}           {\ensuremath{\overline{\Xi}^+}\xspace}
\newcommand{\Xis}          {\ensuremath{\Xi^{\pm}}\xspace}
\newcommand{\Oms}          {\ensuremath{\Omega^{\pm}}\xspace}
\newcommand{\degree}       {\ensuremath{^{\rm o}}\xspace}
\newcommand{\py}           {\ensuremath{\mathrm{PYHIA8~Angantyr}\xspace}}
\newcommand{\hi}           {\ensuremath{\mathrm{HIJING}\xspace}}
\newcommand{\traj}         {\ensuremath{\mathrm{Trajectum}\xspace}}

\begin{titlepage}
\PHyear{2025}       
\PHnumber{126}      
\PHdate{2 June}     

\title{Study of \meanpt and its higher moments, and extraction of the speed of sound in \PbPb collisions with ALICE}
\ShortTitle{Fluctuations of $[\pt]$ and extraction of $c_{\mathrm{s}}^{2}$ in the QGP with ALICE}  
%
\Collaboration{The ALICE Collaboration%
         \thanks{See Appendix~\ref{app:collab} for the list of collaboration
                      members}}
\ShortAuthor{The ALICE Collaboration}      

\begin{abstract}
Ultrarelativistic heavy-ion collisions produce a state of hot and dense strongly interacting QCD matter called quark--gluon plasma (QGP). On an event-by-event basis, the volume of the QGP in ultracentral collisions is mostly constant, while its total entropy can vary significantly due to quantum fluctuations, leading to variations in the temperature of the system. Exploiting this unique feature of ultracentral collisions allows for the interpretation of the correlation of the mean transverse momentum $(\langle p_{\mathrm{T}} \rangle)$ of produced charged hadrons and the number of charged hadrons as a measure for the speed of sound, $c_{\mathrm{s}}$. This speed is related to the rate at which compression waves travel in the QGP and is determined by fitting the relative increase in $\langle p_{\mathrm{T}} \rangle$ with respect to the relative change in the average charged-particle density $(\langle \mathrm{d}N_\mathrm{ch}/\mathrm{d}\eta \rangle)$ measured at midrapidity. This study reports the event-average \meanpt of charged particles as well as the variance, skewness, and kurtosis of the event-by-event transverse momentum per charged particle $([p_{\mathrm{T}}])$ distribution in ultracentral Pb--Pb collisions at a center-of-mass energy of $5.02~\mathrm{TeV}$ per nucleon pair using the ALICE detector. Different centrality estimators based on charged-particle multiplicity or the transverse energy of the event are used to select ultracentral collisions. By ensuring a pseudorapidity gap between the region used to define the centrality and the region used to perform the measurement, the influence of biases and their potential effects on the rise of the mean transverse momentum is tested. The measured $c_{\mathrm{s}}^{2}$ is found to strongly depend on the exploited centrality estimator and ranges between $0.1146 \pm 0.0028 \,\mathrm{(stat.)} \pm 0.0065 \,\mathrm{(syst.)}$ and $ 0.4374 \pm 0.0006 \mathrm{(stat.)} \pm 0.0184 \mathrm{(syst.)}$ in natural units. The self-normalized variance shows a steep decrease towards ultracentral collisions, while the self-normalized skewness variables show a maximum, followed by a fast decrease. These non-Gaussian features are understood in terms of the vanishing of the impact-parameter fluctuations contributing to the event-to-event $[p_{\mathrm{T}}]$ distribution.
\end{abstract}
\end{titlepage}

\setcounter{page}{2} 

\section{Introduction} 
\label{sec:introduction}

It is well established that collisions of heavy ions at ultrarelativistic energies produce a quark--gluon plasma (QGP)~\cite{BRAHMS:2004adc,PHENIX:2004vcz,PHOBOS:2004zne,STAR:2005gfr,Schukraft:2011na,Busza:2018rrf,ALICE:2022wpn}, a state of matter in which quarks and gluons are deconfined and not bound inside hadrons. The QGP formed in a collision undergoes a quick phase of thermalization~\cite{Schlichting:2019abc} before it expands as a relativistic hydrodynamic fluid. The hydrodynamic description of the QGP stands as one of the great successes in developing an effective theory of many-body quantum chromodynamics (QCD) at high temperatures ~\cite{Romatschke:2017ejr,Busza:2018rrf,Bernhard:2019bmu}. As the system expands, both its energy and entropy density decrease and eventually the system undergoes a phase transition as a consequence of which hadrons are formed~\cite{PhysRevC.98.034909,Mazeliauskas:2018irt}. 

It has been suggested that the QGP phase can be studied by measuring the mean transverse momentum $(\meanpt)$ of the produced hadrons in ultracentral \PbPb collisions~\cite{Gardim:2019brr,Nijs:2023bzv,Gardim:2019xjs}. On an event-by-event basis, the volume of the QGP in ultracentral collisions is mostly constant, while the charged-particle multiplicity\,$(\nch)$ can vary significantly~\cite{Gardim:2019brr}. The increase of the charged-particle multiplicity is interpreted as fluctuations in the entropy, which is created early in the collision primarily through interactions of the sea gluons of the colliding nuclei~\cite{VANHOVE1975243}.  As the volume is mostly constant, the corresponding rise in the entropy density $(s)$ leads to higher temperatures $(T)$, as the entropy density is approximately proportional to $T^{3}$ for the QCD equation of state of high temperature deconfined matter~\cite{HotQCD:2014kol}.

The dependence of the pressure, $P$, of the QGP on the energy density, $\epsilon$, is encoded in the corresponding QCD equation of state $P = P (\varepsilon)$~\cite{Braun-Munzinger:2015hba}. The equation of state determines how gradients in the energy density profile give rise to pressure gradients~\cite{Busza:2018rrf}. These gradients of pressure accelerate fluid elements, and facilitate a collective expansion. A fundamental quantity that characterizes the expansion of hot dense matter is the speed of sound, denoted as \cs, which is the speed at which a compression wave travels in a medium. In a relativistic fluid, it is given by $\scs = \mathrm{d}\,P / \mathrm{d}\,\epsilon = \mathrm{d\, ln}\,T / \mathrm{d\, ln}\,s$ ~\cite{Ollitrault:2007du}. Assuming that the increase in the average transverse momentum is solely due to temperature fluctuations, and the charged-particle multiplicity is  proportional to the entropy density~\cite{Gardim:2019xjs,VANHOVE1982138} of the QGP, the speed of sound can be determined experimentally~\cite{CMS:2024sgx,ATLAS:2024jvf} as $\scs = \mathrm{d\, ln}\,\meanpt / \mathrm{d\, ln}\,\nch$. However, this approach does not account for the contribution to \meanpt from radial flow, which is proportional to the inverse size, $1/R$, of the overlap region~\cite{Bozek:2012fw,Bozek:2017elk} --radial flow pushes the \meanpt to higher values with increasing centrality~\cite{ALICE:2018vuu}.

The study of the event-by-event distribution of transverse momentum per charged particle, denoted by $[\pt]$, in ultracentral collisions serves as a tool to probe quantum fluctuations of the initial stage of the collision~\cite{Samanta:2023kfk,Samanta:2023amp}. For collisions with the same \nch, $[\pt]$ fluctuations arise from impact parameter $(b)$ variations and from a quantum nature. Quantum fluctuations originate from the event-to-event positions of the nucleons when colliding and the partonic content of the nucleons~\cite{Broniowski:2009fm,Bozek:2012fw}. At a fixed-impact parameter, $[\pt]$ fluctuations are small and approximately Gaussian distributed but a non-zero skewness is predicted to be driven by event-to-event impact-parameter fluctuations. In particular, Ref.~\cite{Samanta:2023kfk} predicts a rapid increase of the skewness for \nch beyond the \textit{knee} $(N_{\mathrm{ch,knee}})$ marking the rapid decline of the multiplicity distribution for central collisions, followed by a fast decrease. The multiplicity $N_{\mathrm{ch,knee}}$ at the knee is defined as the average multiplicity of collisions at $b=0$.

The ATLAS Collaboration has reported the measurement of the higher-order moments of $[\pt]$ in central collisions~\cite{ATLAS:2024jvf}. In particular, the $Var([\pt])$ features a steep decrease towards the ultracentral collision regime. This striking phenomenon is described in terms of the disappearance of the impact-parameter fluctuations in collisions with the largest multiplicity. Additionally, such observations can only be explained by the presence of a thermalized medium early in the collision, and can serve as a probe of the transport properties of the QGP relying on isotropic expansion instead of anisotropic flow~\cite{Samanta:2023amp}. The ALICE Collaboration previously measured the skewness and kurtosis of the event-by-event mean transverse momentum in wide bins of average charged-particle density $(\avdndeta)$ across different systems~\cite{ALICE:2023tej}. In this article, the study of the higher-order moments of $[\pt]$ is restricted to the ultracentral \PbPb collisions at \fivenn.

This study reports the measurement of the normalized variance $(k_{2})$, normalized skewness $(k_{3})$, standardized skewness $(\gamma_{\langle[\pt]\rangle})$, intensive skewness $(\Gamma_{\langle[\pt]\rangle})$, and standardized kurtosis $(\kappa_{\langle[\pt]\rangle})$ of the event-by-event $[\pt]$ distribution, as well as the event-average \meanpt and the event-average \avdndeta in ultracentral collisions divided into narrow centrality intervals. The data set corresponds to those collisions with the top 0--5\% highest charged-particle multiplicities and top 0--5\% highest transverse energy. In this article, `ultracentral collisions' denotes the 0--0.1\% or smaller fractions of these events with the highest charged-particle multiplicities or transverse energy. A primary purpose of this study is to compare results obtained using charged-particle multiplicity $(N_{\mathrm{ch}})$ centrality estimators with those using transverse-energy $(E_{\mathrm{T}})$ centrality estimators. Furthermore, different kinematic selections on the particles used to define centrality are employed, as it has been found that the definition of the centrality estimator used to measure the \meanpt and \avdndeta can influence the extracted values of \scs~\cite{Nijs:2023bzv,Gardim:2024zvi}. Some of the centrality estimators feature a different choice of pseudorapidity gap with respect to the region used to determine \meanpt as a function of \avdndeta. This is particularly relevant to select high-multiplicity events (ultracentral collisions) with a suppressed contribution of particles from jet fragmentation.

This article is organized as follows. The ALICE experimental setup is described in Sec.~\ref{sec:ExpSetup}, focusing on the detectors which are relevant to the presented measurements. Section~\ref{sec:EvtTrkSel} discusses the analyzed data samples, the event and track-selection criteria, the centrality-estimator definitions, and the analysis techniques to measure the higher-order moments of $[\pt]$ and the \meanpt versus \avdndeta correlation. Section~\ref{sec:EvtTrkSel} also outlines the estimation of systematic uncertainties. The results are presented and discussed in Sec.~\ref{sec:results}, including comparisons to Monte Carlo model predictions. Finally, Sec.~\ref{sec:Conclusions} gives the summary and draws the conclusions.

\section{Experimental setup}
\label{sec:ExpSetup}
A detailed description of the ALICE detector and its performance is provided in Refs.~\cite{ALICE:2008ngc,ALICE:2014sbx}. Relevant detectors for this study include the \VZERO detector~\cite{ALICE:2013axi}, the Inner Tracking System (ITS)~\cite{ALICE:2010tia}, the Time Projection Chamber (TPC)~\cite{Alme:2010ke}, and the Zero Degree Calorimeters (ZDC)~\cite{PUDDU2007397,OPPEDISANO2009206}. 

The \VZERO detector is composed of two scintillator arrays placed along the beam axis ($z$) on each side of the interaction point $(z=0)$: \VZEROA at $z =340~\mathrm{cm}$ and \VZEROC at $z=-90~\mathrm{cm}$. These arrays cover the pseudorapidity regions $2.8<\eta<5.1$ (V0A) and $-3.7<\eta<-1.7$ (V0C). The \VZERO detector provides the minimum bias trigger, which is defined by the requirement of signals in both \VZEROA and \VZEROC detectors in coincidence with a particle bunch crossing corresponding to a beam--beam collision~\cite{ALICE:2013axi}. The \VZERO signals are proportional to the total charge deposited in the scintillators, which correlates with the charged-particle multiplicity in the \VZERO acceptance. The \VZERO detector is also used for centrality estimation~\cite{ALICE:2013hur} and for removing beam induced (beam--gas) background based on timing information. 

The \ITS and \TPC detectors are located within a solenoid that provides a maximum $0.5~\mathrm{T}$ magnetic field parallel to the beam axis. The \ITS is a six-layer silicon detector~\cite{ALICE:2010tia}, surrounding the beam pipe. The two innermost layers comprise the Silicon Pixel Detector (\SPD), located at average distances of 3.9 and 7.6 cm from the beam line with a pseudorapidity coverage of $|\eta|<2$ and $|\eta|<1.4$, respectively. The track segments joining hits in the two \SPD layers are called \textit{tracklets}. The number of tracklets $(\ntracklets)$ is used to estimate the number of primary charged particles produced in the collisions. The Silicon Drift Detector (SDD) comprises the next two layers of the \ITS. In addition to tracking, the \SDD provide charged-particle identification via the measurement of the specific ionization energy loss $(\mathrm{d}E/\mathrm{d}x)$. The \TPC is the main tracking detector, covering the pseudorapidity range $|\eta|<0.9$ with full azimuthal coverage. By measuring drift time, the \TPC provides three-dimensional space-point information for each charged track, with up to 159 space points. Tracks originating from the primary vertex can be reconstructed down to $\pt\sim\mevc{100}$~\cite{ALICE:2014sbx}. Charged-particle multiplicity $(\nch)$ and a proxy for the transverse energy $(\et)$ measured with the \TPC detector are used to estimate the collisions centrality. Transverse energy is quantified as the summed transverse mass $(m_{\mathrm{T}}=\sqrt{\pt^{2}+m_{\pi}^{2}})$, assuming the pion mass for all particles. 

The ZDC measures the energy of the spectator nucleons in the forward direction, providing a direct estimate of the average number of participating nucleons $(\langle N_{\mathrm{part}} \rangle)$ in the collisions. The $\langle N_{\mathrm{part}} \rangle$ calculation is valid for central collisions (0--5\%) where the contribution from nuclear fragments that escape detection by the \ZDC is negligible~\cite{ALICE:2013hur}. The neutron (ZNC and ZNA) calorimeters are placed at zero degrees with respect to the LHC beam axis to detect forward going neutral particles at pseudorapidities $|\eta|>8.8$, while the proton (ZPC and ZPA) calorimeters are located externally to the outgoing beam vacuum tube. In this study, the \ZDC detector is only used to estimate the centrality dependent $\langle \Npart \rangle$~\cite{ALICE:2013hur}. 

\section{Analysis procedure}

\subsection*{Event and track selection}
\label{sec:EvtTrkSel}

The present study uses data from \PbPb collisions at \fivenn collected during the Run 2 data-taking period of the LHC in 2018. The primary-vertex position is reconstructed using information from both the \TPC and \ITS detectors. A $\pm10~\mathrm{cm}$ selection is applied to the primary-vertex position along the beam axis to ensure uniform pseudorapidity coverage in the \SPD and \TPC detectors at midrapidity. The total number of minimum bias collisions analyzed after event and vertex selections amounts to about 193 million.

This analysis uses primary charged particles, which are defined as charged particles produced directly in the collision with a mean proper lifetime $\tau$ that is larger than $1\,\mathrm{cm}/c$, or from decays of particles produced at the interaction point with $\tau$ shorter than $1\,\mathrm{cm}/c$, excluding daughters from long-lived weakly decaying hadrons and particles produced in interactions with the detector material~\cite{ALICE:2017hcy}. Tracks of primary charged particles are reconstructed using the combined information from the \ITS and \TPC detectors. The track-selection criteria are the same as the ones used in previous studies~\cite{ALICE:2018vuu}, and yield the best track quality and minimal contamination from secondary particles. The reconstructed tracks are required to have a minimum ratio between crossed rows and reconstructed space points in the \TPC of 0.8. The fit quality for the \ITS and \TPC track points must satisfy $\chi^{2}_{\mathrm{ITS}}/N_{\mathrm{hits}} < 36$ and $\chi^{2}_{\mathrm{TPC}}/N_{\mathrm{clusters}} < 4$, where $N_{\mathrm{hits}}$ and $N_{\mathrm{clusters}}$ are the number of hits in the \ITS and the number of reconstructed space points in the \TPC associated to a track, respectively. To limit the contamination from secondary particles, the distance-of-closest approach (DCA) to the primary vertex in the transverse plane has to satisfy the \pt-dependent selection: $|\mathrm{DCA}_{xy}|<A+B\cdot\pt^{C}$, with $A=0.0182~\mathrm{cm}$, $B=0.035~\mathrm{cm}$, and $C=-1.01$. The \pt is the numerical value of the transverse momentum in units of $\mathrm{GeV}/c$. A 2 cm selection is also applied to the $\mathrm{DCA}$ along the $z$ axis. Finally, primary charged particles are measured in the kinematic range $|\eta|\leq0.8$ and $0.15\leq \pt < 50~\mathrm{GeV}/c$.

\subsection*{Selecting ultracentral collisions}
\label{sec:centrality_selection}

A primary objective of this analysis is to investigate the recently reported dependence of the measured \cs on the acceptance, kinematic selections, and the observable used to determine the collision centrality~\cite{Nijs:2023bzv,Gardim:2024zvi}. Table~\ref{tab:centrality_definition} summarizes the different centrality estimators, including the kinematic selections on particles used for centrality estimation and for the measurement of \meanpt and \avdndeta. Centrality estimators based on the number of \SPD tracklets include particles with transverse momenta starting from approximately $0.03~\mathrm{GeV}/c$ and have no upper \pt limit. In contrast, the centrality estimators using the number of charged particles reconstructed with the \TPC are constrained to $0.15\leq \pt <50~\mathrm{GeV}/c$. This also applies to the \et-based centrality estimators.

Significant autocorrelation effects are expected when the pseudorapidity intervals used for centrality estimation and for \meanpt and \avdndeta measurement completely overlap. This occurs when event activity for centrality assessment is quantified in $|\eta|\leq0.8$, and \meanpt and \avdndeta are measured within the same pseudorapidity window, as represented by labels I, III, and V in Table~\ref{tab:centrality_definition}. Specifically, a multiplicity bias is expected with \nch-based estimators, and an energy bias is expected with \et-based estimators.

A pseudorapidity gap is introduced between the centrality estimation region and the region used for \meanpt and \avdndeta measurement. This is particularly important for suppressing the effects of particles from jet fragmentation. The fragmentation of jets into charged particles with intermediate to high \pt can increase both \meanpt and \avdndeta, which may not necessarily reflect an increase in the entropy density of the QGP. 

A pseudorapidity gap is introduced for centrality estimators based on the \nch in \TPC (II), \et in \TPC (IV), \ntracklets in \SPD (VI and VIII), and \nch in \VZERO (IX), as defined in Table~\ref{tab:centrality_definition}. Notably, estimators labeled VIII and IX allow for the investigation of \meanpt and \avdndeta dependence with a wider pseudorapidity gap. 

This analysis also includes a case where the pseudorapidity region for centrality estimation is adjacent to the region used for \meanpt and \avdndeta measurement. This case utilizes the \ntracklets in the \SPD detector and is labeled as VII.

\begin{table}[!ht]
\caption[]{The columns, from left to right, present: the observable used for centrality estimation, the label identifying each estimator in the figures, the pseudorapidity interval used to measure event activity for centrality classification, the pseudorapidity interval used to measure \meanpt and \avdndeta, and the minimum pseudorapidity gap between the centrality estimation region and the region to measure \meanpt and \avdndeta. The \pt selections for charged tracks and tracklets are given in the text.}
\centering
\begin{tabular}{cccccc}
 \toprule
 Observable & Label & Centrality estimation & $\meanpt$ and $\avdndeta$ & Minimum $|\Delta\eta|$\\
 \midrule
 \multirow{2}{*}{\nch in \TPC} & I & $|\eta|\leq0.8$ & $|\eta|\leq0.8$ & 0 \\
                               & II & $0.5\leq|\eta|<0.8$ & $|\eta|\leq0.3$ & 0.2 \\
 \midrule
 \multirow{2}{*}{\et in \TPC} & III & $|\eta|\leq0.8$ & $|\eta|\leq0.8$ & 0 \\
                              & IV & $0.5\leq|\eta|<0.8$ & $|\eta|\leq0.3$ & 0.2 \\
 \midrule
 \multirow{3}{*}{\ntracklets in \SPD} & V & $|\eta|\leq0.8$ & $|\eta|\leq0.8$ & 0 \\
                               & VI & $0.5\leq|\eta|<0.8$ & $|\eta|\leq0.3$ & 0.2 \\
                               & VII & $0.3<|\eta|<0.6$ & $|\eta|\leq0.3$ & 0 \\
                               & VIII & $0.7\leq|\eta|<1$ & $|\eta|\leq0.3$ & 0.4 \\
 \midrule
 \nch in \VZERO & IX & $-3.7<\eta<-1.7$ and $2.8<\eta<5.1$ & $|\eta|\leq0.8$ & 0.9 \\
 \bottomrule
\end{tabular}
\label{tab:centrality_definition}
\end{table}

Previous ALICE publications employed a phenomenological approach to extract the average number of participating nucleons, \avNpart, relying on a Glauber Monte Carlo calculation convoluted with a negative binomial distribution (NBD) model for particle production to fit the \VZERO amplitude distribution~\cite{ALICE:2013hur}. In this analysis, a data-driven method is employed to measure the \avNpart in the 0--5\% centrality interval.

\begin{figure}[!ht]
	\centering
	\hspace{0cm}
    \includegraphics[width=0.95\textwidth]{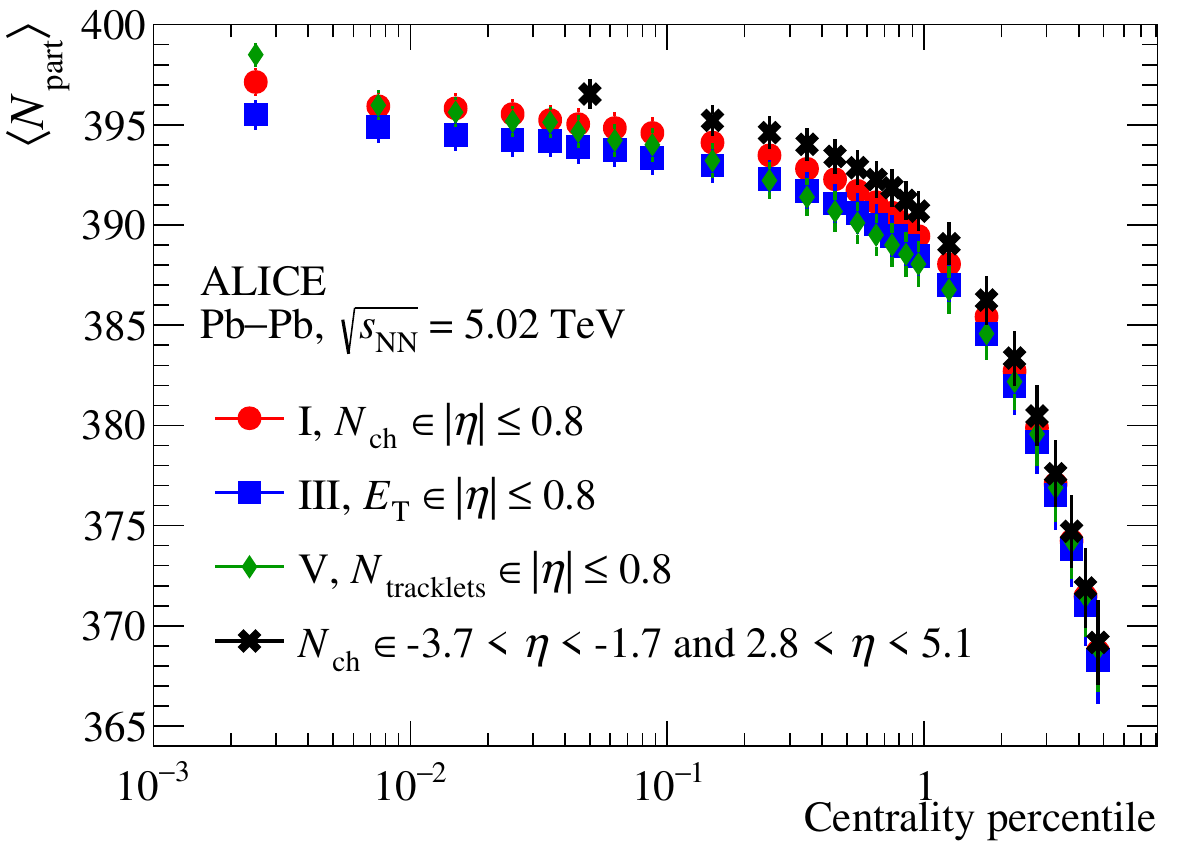}
	\hspace{0cm}
	\caption{Average number of participating nucleons $(\avNpart)$ as a function of centrality percentile in \PbPb collisions at \fivenn. Data points are shown for centrality estimators based on \nch, \et, \ntracklets within $|\eta|\leq 0.8$, and \nch within $-3.7<\eta<-1.7$ and $2.8 < \eta <5.1$. Uncertainty bars represent the sum of statistical and systematic uncertainties, with the latter being the dominant source. The systematic uncertainty is determined by varying the acceptance correction factors within their uncertainties and assigning the maximum deviation from the nominal \Npart value as the systematic uncertainty.}
	\label{fig:Npart_vs_cent}
\end{figure}

Figure~\ref{fig:Npart_vs_cent} illustrates the centrality dependence of \avNpart, computed from the average number of spectator nucleons reaching the \ZDC detector~\cite{ALICE:2013hur}. The \avNpart is determined using the following equation

\begin{equation}
\avNpart = 2\mathrm{A} - \Bigg( \frac{\langle E_{\mathrm{ZNC}} \rangle}{\alpha_{\mathrm{ZNC}}} + \frac{\langle E_{\mathrm{ZNA}} \rangle}{\alpha_{\mathrm{ZNA}}} + \frac{\langle E_{\mathrm{ZPC}} \rangle}{\alpha_{\mathrm{ZPC}}} +\frac{\langle E_{\mathrm{ZPA}} \rangle}{\alpha_{\mathrm{ZPA}}}\Bigg) /E_{\mathrm{A}}~,
\end{equation}

where $\mathrm{A} = 208$  is the mass number of the Pb nucleus, $E_{\mathrm{A}}=2.51~\mathrm{TeV}$ is the beam energy per nucleon, $\langle E_{\mathrm{ZNC}} \rangle, \langle E_{\mathrm{ZNA}} \rangle, \langle E_{\mathrm{ZPC}}\rangle$, and $\langle E_{\mathrm{ZPA}}\rangle$ represent the neutron and proton energies deposited in the neutron and proton calorimeters on each side of the interaction point, and $\alpha_{\mathrm{ZNC}}=0.933\pm 0.0165, \alpha_{\mathrm{ZNA}}=0.931\pm 0.0164, \alpha_{\mathrm{ZPC}}=0.5\pm0.05$, and $\alpha_{\mathrm{ZPA}}=0.52\pm0.07$ are the corresponding corrections for detection efficiency and acceptance, calculated with Monte Carlo simulated events~\cite{ALICE:2024vpj}. The uncertainties in the proton correction factors, which encompass variations in beam optics during \PbPb data taking in 2018, are included in the \avNpart estimation. 

Figure~\ref{fig:Npart_vs_cent} presents results for centrality estimators using \nch, \et, \ntracklets within $|\eta| \leq 0.8$, and \nch within $-3.7<\eta<-1.7$ and $2.8 < \eta <5.1$. Similar results are obtained for the other centrality estimators. The data demonstrate a consistent trend across all estimators, regardless of whether charged-particle multiplicity or transverse energy is used for event classification. The \avNpart increases rapidly from the 4.5--5\% to the 0.9--1\% centrality interval and then exhibits a slight saturation for the most central collisions. The relative increase of \avNpart in the 0--0.005\% centrality interval compared to the 0.9--1\% interval is approximately 1\%, suggesting that the volume of the QGP remains relatively constant in the ultracentral-collision limit. 

The \avNpart values in the 0--0.1\% centrality range, determined using the \et centrality estimator, are systematically lower compared to those obtained with \nch centrality estimators, indicating distinct selection biases. Specifically, collisions characterized by lower \avNpart tend to exhibit lower \nch at midrapidity and greater impact-parameter fluctuations. The \nch-based centrality estimators generally select higher \avNpart values for the same centrality interval than the \et-based estimator. These selection biases are described in detail in Sec.~\ref{sec:results} when examining the evolution of \meanpt as a function of the centrality estimators. Conversely, the \ntracklets-based centrality estimator yields the largest \avNpart increase from 1\% to 0\% centrality. Finally, the \VZERO-based centrality estimator employs coarser binning for the most ultracentral collisions.

The estimation of \avNpart with the \VZERO-based centrality estimator yields a value of $388$ for the 0--5\% centrality class, which agrees within $1\%$ with a calculation using a NBD-Glauber fit to the \VZERO amplitude distribution~\cite{ALICE:2015juo}.  It is interesting to note that \avNpart does not reach the asymptotic value of 416. This is expected, as previous calculations show the radius of the overlap region saturates at around 6 fm~\cite{Gardim:2019brr}, which is below the Pb nucleus radius of 6.7 fm. Therefore, reaching the asymptotic value is not possible, as the nuclei never fully overlap in the ultracentral region explored.
 
\subsection*{Measuring $\mathbf{\langle \textit{p}_{T} \rangle}$, $\mathbf{\langle d\textit{N}_{ch}/d\eta \rangle}$, and the higher-order moments of $\mathbf{[\textit{p}_{T}]}$}
\label{sec:measuring_meanpT}

The speed of sound is extracted from a fit to the correlation between the normalized event-average transverse momentum, $\ptnorm = \normmeanpt$, and the normalized event-average charged-particle density $\dndetanorm= \normdndeta$, where the normalization constants are measured in the 0--5\% centrality range. Both quantities are derived from the centrality-dependent transverse-momentum spectra fully corrected for the acceptance, tracking inefficiency, and secondary-particle contamination. The spectra are measured and corrected using standard methods~\cite{ALICE:2018vuu}, which involve employing HIJING event simulations~\cite{PhysRevD.44.3501}. The generated particles are subsequently propagated through a simulation of the ALICE detector using the GEANT 3 transport code~\cite{Geant3}. The simulated particles are reconstructed using the same algorithms as for the data. The corrections are determined using a high-multiplicity sample, specifically collisions in the 0--5\% centrality interval. The tracking-inefficiency correction accounts for the particle composition of the charged-hadron spectrum. The transverse-momentum dependent fractions of charged pions, kaons, protons, and sigma baryons are used to refine the Monte Carlo-based tracking inefficiency correction~\cite{ALICE:2018vuu}. The residual contamination from secondary particles (products of weak decays and particles produced from interactions with the detector material) is estimated using a data-driven approach based on a multi-template fit of the data $\mathrm{DCA}_{xy}$ distributions in transverse-momentum intervals. The $\mathrm{DCA}_{xy}$  distributions are fitted with three Monte Carlo templates representing the contribution from primary and secondary particles, with the latter originating from weak decays or from interactions with the detector material. The fraction of secondary particles amounts to 12\% at $\pt=0.15~\mathrm{GeV}/c$ and decreases asymptotically to about 2\% at $\pt=3.5~\mathrm{GeV}/c$. 

The \meanpt and \avdndeta are derived from the \pt spectra in the interval $0\leq\pt\leq 10~\mathrm{GeV}/c$. Prior to the calculation, an extrapolation procedure is applied to estimate the unmeasured yield in the interval between $0\leq\pt< 0.15~\mathrm{GeV}/c$. The extrapolation procedure closely follows that described in Refs.~\cite{ALICE:2019hno,ALICE:2020nkc}, where the transverse-momentum spectra are fitted with a Boltzmann-Gibbs Blast-Wave model~\cite{Schnedermann:1993ws} in the interval between $0.15\leq\pt\leq 1.5~\mathrm{GeV}/c$. The fit range is selected based on the $\chi^{2}/\mathrm{ndf}$ criterion. The extrapolated \pt-integrated yield amounts to approximately 9\% of the yield in the $0\leq\pt\leq 10~\mathrm{GeV}/c$ interval. 

The statistical uncertainty on \meanpt and \avdndeta is calculated by shifting each data point by a fraction of its statistical uncertainty. The fraction is randomly drawn from a Gaussian distribution with a standard deviation of 1, and new values of integrated yields and mean transverse momenta are calculated. The procedure is repeated 1000 times, and the standard deviations of \meanpt and \avdndeta are used as the statistical uncertainties.

The reported higher-order moments of $[\pt]$, defined in~\crefrange{eq:moments}{eq:cumulants}, are the normalized variance $(\ktwnorm = \ktwonorm)$, skewness $(\kthnorm = \kthreenorm)$, standardized skewness $(\stdsknorm = \stdskewnorm)$, intensive skewness $(\intsknorm = \intskewnorm)$, and kurtosis $(\kunorm = \kurnorm)$ as a function of (\dndetanorm). The self-normalized quantities provide precise measurements of the relative variations in ultracentral collisions with respect to their values in the 0--5\% centrality class since the common uncertainties between the numerator and denominator cancel out.

The cumulants are genuine correlations unbiased by contributions from lower-order correlations and are related to the properties of the distribution, such as the variance, skewness, and kurtosis. The cumulants are constructed from the $\pt$ correlations, $\mptk{k}$, given by~\cite{Nielsen:2023znu}
\begin{align}
    \mptk{k} = \frac{\displaystyle\sum_{i_1\neq\ldots\neq i_k}w_{i_1}\ldots w_{i_k}p_{\mathrm{T},i_1} \ldots p_{\mathrm{T},i_k} }{\displaystyle\sum_{i_1\neq\ldots\neq i_k}w_{i_1}\ldots w_{i_k}},
\end{align}
where the index runs over distinct $k$-particle tuplets, $i_1\neq\ldots\neq i_k$, and $w_i$ are particle weights to correct for non-uniform efficiencies of the detectors. The second-, third-, and fourth-order $\pt$-cumulants are
\begin{align}
\label{eq:moments}
c_2 &= \mmptk{2}-\mmpt^2,\\
c_3 &= \mmptk{3}-3\mmptk{2}\mmpt+2\mmpt^3,\\
c_4 &= \mmptk{4}-4\mmptk{3}\mmpt-3\mmptk{2}^2+12\mmptk{2}\mmpt^2-6\mmpt^4.
\end{align}
The cumulants are then converted to their normalized, dimensionless form
\begin{align}
k_2 = \frac{c_2}{\mmptk{2}}, \quad k_3 = \frac{c_3}{\mmptk{3}}, \quad \stdskew = \frac{c_3}{c_2^{3/2}}, \quad \intskew = \frac{c_3\cdot \mmpt}{c_2^2}, \quad \kur = \frac{c_4+3c_2^2}{c_2^2}.
\label{eq:cumulants}
\end{align}

The higher-order cumulants are measured at midrapidity ($|\eta|<0.8$) and within $0.2<\pt<3$ GeV/\textit{c} as a function of the event-by-event charged-particle multiplicity density and transverse energy in the same kinematic phase space. The statistical uncertainty is estimated using standard procedures~\cite{ALICE:2024fcv}, which employ the bootstrap method of random sampling with replacement~\cite{e89fac9c-03d7-3e22-aa30-08f5596f8fce}.

\subsection*{Systematic uncertainties}
\label{sec:systematic_uncertainties}

This section describes the calculation of the total systematic uncertainty on the average transverse momentum, the event-by-event higher-order mean transverse-momentum cumulants, and the average charged-particle density. Two sources of systematic uncertainty are considered. 

The first is due to the used vertex and track selections. The effect of selecting events based on the vertex position is studied by comparing the default results to the fully corrected \ptnorm and \dndetanorm obtained with an alternative vertex selection corresponding to $\pm5~\mathrm{cm}$ and for the high-order cumulants corresponding to $\pm 7$, and $\pm9~\mathrm{cm}$. The average relative systematic uncertainty is 0.02\% for \ptnorm and effectively zero for \dndetanorm in the 0--5\% centrality range. However, in the higher-order mean transverse-momentum cumulants analysis, it was found to be statistically insignificant based on the Barlow criterion~\cite{Barlow:2002yb}. The systematic uncertainty due to the track-selection criteria is investigated by varying the selections employed on the tracks. In particular, the minimum ratio between crossed rows and reconstructed space points in the \TPC is shifted to 0.7 and 0.9 (the nominal is 0.8). The $\chi^{2}_{\mathrm{ITS}}/N_{\mathrm{hits}}$ is set to 25 and 49 (the nominal is 36), while the $\chi^{2}_{\mathrm{TPC}}/N_{\mathrm{clusters}}$ is shifted to 3 and 5 (the nominal is 4). The $\mathrm{DCA}_{z}$ selection along the beam axis is also varied to 1 and 5 cm (the nominal is 2 cm). The relative systematic uncertainty for \ptnorm is 0.21\% in the most central collisions and decreases with decreasing centrality. For the higher-order cumulants, the quality of the reconstructed tracks is varied by increasing the number of TPC space points from a default of 70 to 80 and 90, which leads to less than 0.5\% variation for the results based on midrapidity centrality estimators and less than a 3\% variation for the results based on forward centrality estimator. Additionally, a different track type is considered, which includes additional tracks without hits in the innermost layer of the ITS to recover a uniform distribution as a function of  $\varphi$ and $\eta$. This leads to a negligible variation for midrapidity centrality estimators and around 3\% for the forward centrality estimator. Finally, the variations of the $\mathrm{DCA}$ in the longitudinal and transverse planes lead to differences $<1\%$ in the midrapidity-based cases and $<2\%$ and $<3\%$, respectively, in the forward-based cases. The systematic uncertainty is quantified as: $\mathrm{Unc} = 1 - (X_{\mathrm{var}} / X_{\mathrm{nom}})/(X^{\mathrm{0-5\%}}_{\mathrm{var}} / X^{\mathrm{0-5\%}}_{\mathrm{nom}})$, where $X_{\mathrm{nom}}$ represents the nominal observable (average transverse momentum or higher-order cumulants), and $X_{\mathrm{var}}$ is the value of the same observable for a particular variation. 
The second source of systematic uncertainty, which is only considered for the event-averaged mean transverse momentum, is the choice of the Boltzmann-Gibbs Blast-Wave model to fit the spectra during the extrapolation procedure. This is quantified by measuring the \ptnorm and \dndetanorm using alternative fit functions: the Lévy Tsallis~\cite{Wilk:1999dr}, and Hagedorn~\cite{Hagedorn:1983wk} parameterizations. The maximum \ptnorm and \dndetanorm deviation with respect to the results from using the nominal fit function is assigned as the systematic uncertainty. 

Finally, the total systematic uncertainty is given by the sum in quadrature of the different sources of systematic uncertainty. The dominant source of systematic uncertainty comes from the vertex and track selections. The total relative systematic uncertainty on the \ptnorm is about 0.23\% for the most central collisions and decreases to about 0.06\% for collisions in the 4.5--5\% centrality range. The total systematic uncertainty on the \dndetanorm is negligible.

\section{Results and discussion}
\label{sec:results}

\begin{figure}[!ht]
	\centering
	\hspace{0cm}
    \includegraphics[width=0.95\textwidth]{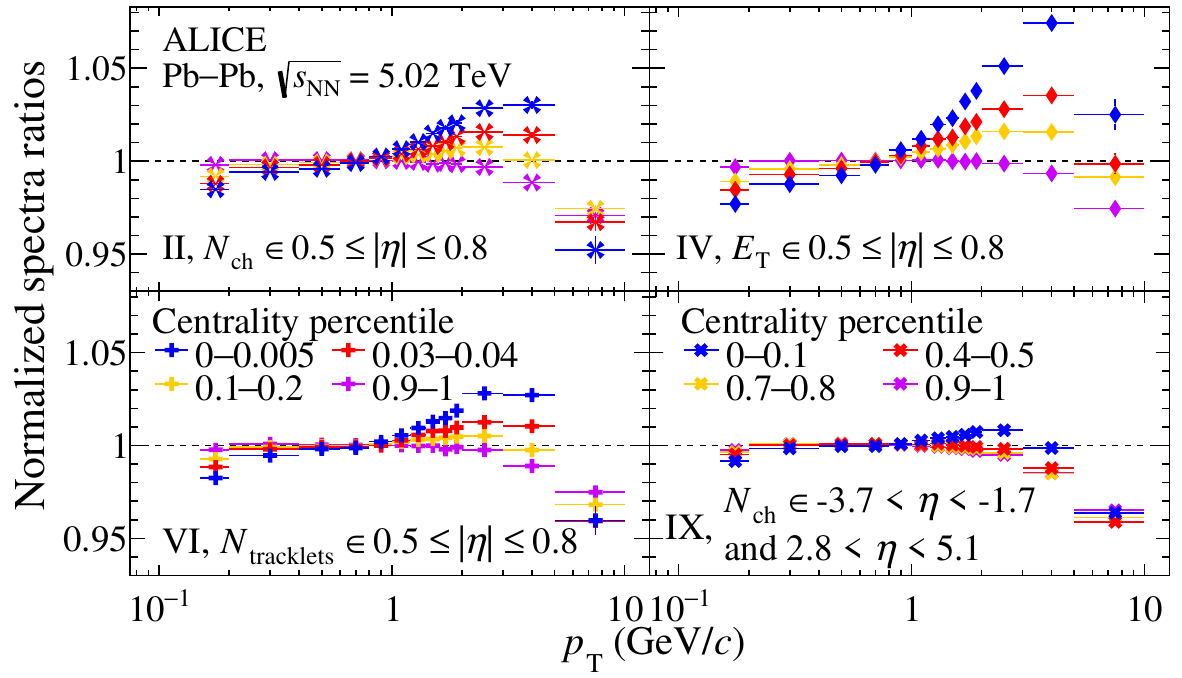}
	\hspace{0cm}
	\caption{Normalized spectra ratios as a function of the transverse momentum in \PbPb collisions at \fivenn. Results are shown for centrality estimators based on \nch (top left), \et (top right), \ntracklets (bottom left) within $0.5 \leq |\eta| \leq 0.8$, and \nch within $-3.7<\eta<-1.7$ and $2.8 < \eta <5.1$ (bottom right). Each panel displays normalized ratios for selected centrality classes. The centrality percentile legend in the bottom left panel applies to midrapidity estimators, while the legend in the bottom right panel applies to the forward estimator. Error bars represent statistical uncertainties. Systematic uncertainties, which are largely canceled due to their common origin in both the numerator and denominator, are not shown.}
	\label{fig:spectra_ratios}
\end{figure}

Figure~\ref{fig:spectra_ratios} shows the ratios of normalized transverse-momentum spectra for the most central collisions using the following centrality estimators: \ntracklets, \nch, \et $\in 0.5 \leq |\eta| \leq 0.8$, and \nch $\in -3.7<\eta<-1.7$ and $2.8 < \eta <5.1$. The normalized ratios are defined as

\begin{equation}
    \frac{(\mathrm{d}^{2}N/\langle \mathrm{d}N_{\mathrm{ch}}/\mathrm{d}\eta \rangle \mathrm{d}\eta\mathrm{d}\pt)^{\mathrm{Centrality~ percentile}}}{(\mathrm{d}^{2}N/\langle \mathrm{d}N_{\mathrm{ch}}/\mathrm{d}\eta \rangle \mathrm{d}\eta\mathrm{d}\pt)^{\mathrm{0-5\%}}}~.
\end{equation}

The normalized ratios are shown for the centrality estimators with pseudorapidity gap between the region to estimate the collision centrality and the region to measure the \pt spectra. The charged-particle multiplicity based centrality estimators with the \SPD and \TPC detectors show a yield depletion for $\pt \lesssim 1 ~\mathrm{GeV}/c$ and a pronounced enhancement in the $1<\pt<6~\mathrm{GeV}/c$ interval with a maximum at $\pt \approx 4~\mathrm{GeV}/c$ for the most central collisions. This observation is reminiscent of radial flow~\cite{ALICE:2015dtd,ALICE:2019hno}. The radial-flow effects are the strongest for the most central collisions and diminish towards less central events. Furthermore, the normalized ratios in the 6--10$~\mathrm{GeV}/c$ transverse-momentum interval decrease, which suggests that any increase of the \meanpt for ultracentral collisions with respect to the average transverse momentum in the reference class (0--5\%) is primarily attributed to the entropy fluctuations and the hydrodynamic expansion of the QGP rather than to the effects of jet fragmentation~\cite{Gardim:2019brr,Gardim:2019xjs}. The normalized \pt-spectra ratios for events selected with the \VZERO detector (bottom right panel in Fig.~\ref{fig:spectra_ratios}) show similar trends although the height of the radial-flow bump is considerably smaller compared to the midrapidity estimators. The spectra ratios with the \et-based centrality estimator (top right panel in Fig.~\ref{fig:spectra_ratios}) also show a yield depletion at low transverse momentum, however they show a sharp ratio increase above $\pt=1~\mathrm{GeV}/c$ for the most central bin, and do not go back below 1 for $\pt \gtrsim 5~\mathrm{GeV}/c$. This observation suggests a tight short-range correlation between the activity in the centrality region and the region where the \pt spectrum is measured, leading to a \pt bias expected to give a higher \meanpt compared with the one found using the \nch-based centrality estimators.

\begin{figure}[!ht]
	\centering
	\hspace{0cm}
    \includegraphics[width=0.95\textwidth]{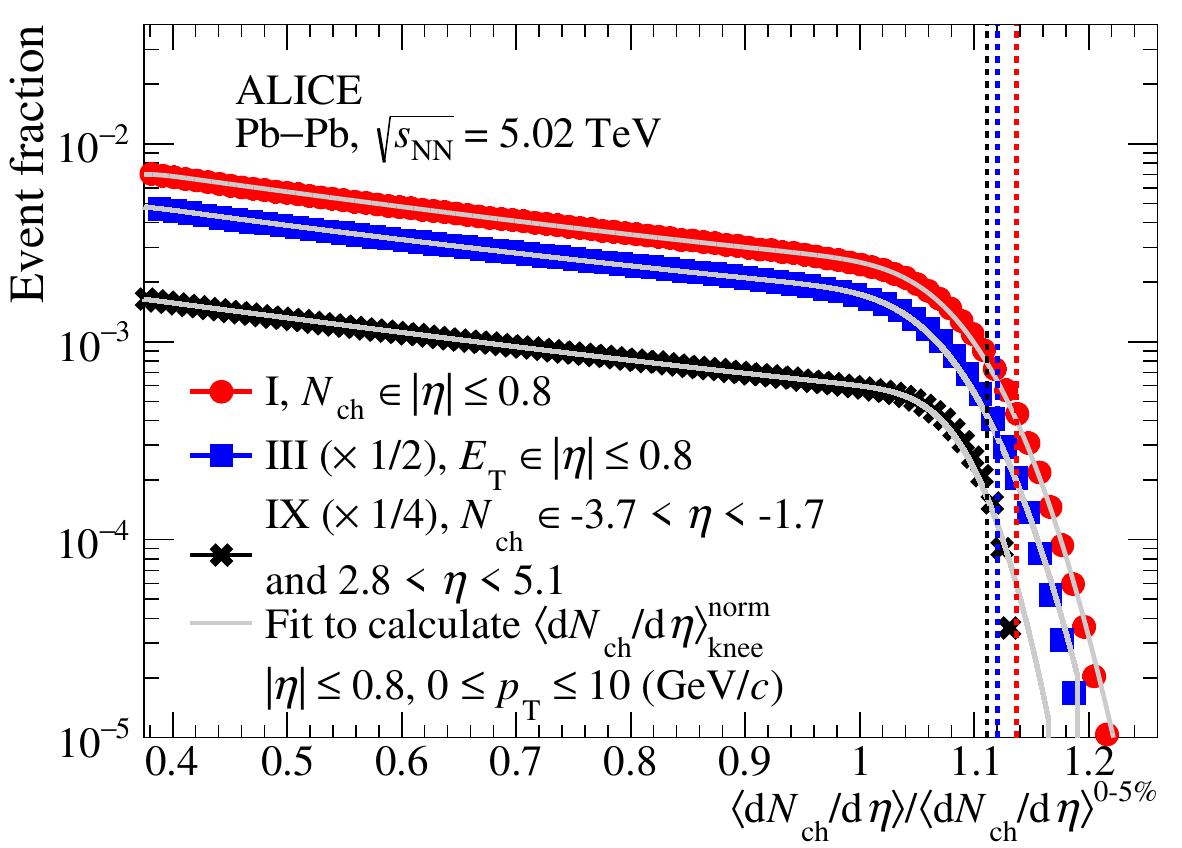}
	\hspace{0cm}
	\caption{Event fraction distribution as a function of the normalized charged-particle density in \PbPb collisions at \fivenn. Centrality classification is based on \nch (red circles) and \et (blue squares) in $|\eta|\leq0.8$, and on forward \nch for the \VZERO (black crosses). \dndetanorm is derived from the extrapolated spectra in $|\eta|\leq0.8$. Gray curves represent fits using the model from Ref.~\cite{Das:2017ned}. The positions of \dndetanormknee, indicated by dashed vertical lines, are 1.137, 1.120, and 1.111 for the \nch-, \et-, and \VZERO-based centrality estimators, respectively.}
	\label{fig:event_fraction}
\end{figure}

Measured charged-particle multiplicity distributions from central collisions are well described by a Gaussian distribution at fixed-impact parameter $(b)$~\cite{Das:2017ned}. The charged-particle density at the \textit{knee} is denoted by $\avdndeta_{\mathrm{knee}}$, and it is defined as the mean value of the charged-particle density distribution for collisions with $b=0$, while the standard deviation of this distribution is represented by $\sigma_{\mathrm{knee}}$~\cite{Das:2017ned}. The mean and standard deviation at the knee are determined by fitting the \dndetanorm-dependent event fraction distribution with a model for the multiplicity distribution for fixed-impact parameter. Importantly, this model does not rely on the concept of participant nucleons or any microscopic model of the collision. The \dndetanorm and $\sigma/\sigma^{0-5\%}$ at the knee are denoted by \dndetanormknee and \sigmanormknee, respectively. The event fraction distribution, $P(\dndetanorm)$, is modeled by the integral of $P(\dndetanorm | b)$ over all values of $b$, where $P(\dndetanorm | b)$ is the probability of \dndetanorm for fixed-impact parameter given by a Gaussian distribution. Each Gaussian is characterized by the normalized mean, $\overline{\avdndeta^{\mathrm{norm}}}(b)$ and the normalized standard deviation, $\sigma^{\mathrm{norm}}(b)$. The employed parametric forms are $\overline{\avdndeta^{\mathrm{norm}}}(b)=\dndetanormknee \, \mathrm{exp}(-a_1 b - a_2 b^{2} - a_3 b^{3})$ and $\sigma^{\mathrm{norm}}(b) = \sigmanormknee$, where $a_1, a_2$, and $a_3$ are free parameters. Ref.~\cite{Das:2017ned} proposes an scenario where $\sigma^{\mathrm{norm}}(b)\propto \overline{\avdndeta^{\mathrm{norm}}}(b)$, which is more suitable to describe the event fraction distribution of central and semicentral collisions. Since, this study focuses on ultracentral collisions, using $\sigma^{\mathrm{norm}}(b) = \sigmanormknee$ describes well the event fraction distribution of central collisions. Figure~\ref{fig:event_fraction} illustrates the event fraction distribution as a function of \dndetanorm derived from collisions selected with the \nch-, \et-, and \VZERO-based centrality estimators. For midrapidity estimators, the region for measuring \dndetanorm completely overlaps with the region for assessing collision centrality. This means both quantities are determined within $|\eta| \leq 0.8$. In Fig.~\ref{fig:event_fraction}, dashed vertical lines indicate the position of \dndetanormknee. Table~\ref{tab:knee_sigma0} provides a list of \sigmanormknee and \dndetanormknee values for all the studied centrality estimators. Fits to the data yielded $\chi^2/\mathrm{ndf}$ values equal to 1.232, 0.538, and 0.028 for the \nch-, \et-, and \VZERO-based centrality estimators, respectively.

\begin{table}[!ht]
\caption[]{Values of \sigmanormknee and \dndetanormknee obtained from fitting the event fraction distribution shown in Fig.~\ref{fig:event_fraction}. The fit parameters are given for all the centrality estimators. The last column lists the values of \scs for all the centrality estimators in natural units. These values are obtained from a fit to the correlation between \ptnorm and \dndetanorm using Eq.~\ref{eq:fitmeanpt}. The definition of each estimator is given in Table~\ref{tab:centrality_definition}.}
\centering
\begin{tabular}{ccccc}
 \toprule
 Label & \dndetanormknee & \sigmanormknee & Speed of sound $(\scs)$ \\
 \midrule
 I & 1.1370 $\pm$ 0.0004 (stat.) & 0.0348  $\pm$ 0.0005 (stat.) & 0.1369 $\pm$ 0.0007 (stat.) $\pm$ 0.0015 (syst.)\\
 II & 1.1040 $\pm$ 0.0021 (stat.) & 0.0202 $\pm$ 0.0006 (stat.) & 0.1795 $\pm$ 0.0018 (stat.) $\pm$ 0.0083 (syst.)\\
 \midrule
 III & 1.1200 $\pm$ 0.0767 (stat.) & 0.0359 $\pm$ 0.0033 (stat.) & 0.4374 $\pm$ 0.0006 (stat.) $\pm$ 0.0184 (syst.)\\
 IV & 1.1010  $\pm$ 0.0131 (stat.) & 0.0201 $\pm$ 0.0006 (stat.) & 0.3058 $\pm$ 0.0015 (stat.) $\pm$ 0.0143 (syst.)\\
 \midrule
 V & 1.1450 $\pm$ 0.0001 (stat.) & 0.0268 $\pm$ 0.0006 (stat.) & 0.1773 $\pm$ 0.0013 (stat.) $\pm$ 0.0066 (syst.)\\
 VI & 1.1090 $\pm$ 0.0006 (stat.) & 0.0185 $\pm$ 0.0012 (stat.) & 0.1873 $\pm$ 0.0025 (stat.) $\pm$ 0.0143 (syst.)\\
 VII & 1.1120 $\pm$ 0.0026 (stat.) & 0.0183 $\pm$ 0.0023 (stat.)& 0.2083 $\pm$ 0.0024 (stat.) $\pm$ 0.0249 (syst.)\\
 VIII & 1.1248 $\pm$ 0.0169 (stat.) & 0.0227 $\pm$ 0.0020 (stat.) & 0.1473 $\pm$ 0.0023 (stat.) $\pm$ 0.0119 (syst.)\\
 \midrule
 IX & 1.1144 $\pm$ 0.0024 (stat.) & 0.0186 $\pm$ 0.0023 (stat.) & 0.1146 $\pm$ 0.0028 (stat.) $\pm$ 0.0065 (syst.)\\
 \bottomrule
\end{tabular}
\label{tab:knee_sigma0}
\end{table}

The speed of sound, \scs is extracted by fitting the \ptnorm versus \dndetanorm correlation to the parameterization proposed in Ref.~\cite{Gardim:2019brr}, based on the relation $\meanpt \propto s^{\scs}$ with $s$ representing the entropy density

\begin{equation}
    \ptnorm = \Bigg(\frac{\dndetanorm}{f(\dndetanorm, \dndetanormknee, \sigmanormknee)}\Bigg)^{\scs},
\label{eq:fitmeanpt}
\end{equation}

with

\begin{equation} 
\label{eq:ffunction}
\begin{split}
f(\dndetanorm, \dndetanorm_{\mathrm{knee}}, \sigma^{\mathrm{norm}}_{\mathrm{knee}}) &= \dndetanorm - \\
 & \sigma^{\mathrm{norm}}_{\mathrm{knee}}\sqrt{\frac{2}{\pi}}\frac{\mathrm{exp}\Big(-\frac{(\dndetanorm - \dndetanorm_{\mathrm{knee}})^{2}}{2 (\sigmanormknee)^{2} }\Big)}{\mathrm{erfc}\Big(\frac{\dndetanorm - \dndetanorm_{\mathrm{knee}}}{\sqrt{2}\,\sigmanormknee}\Big)}.
\end{split}
\end{equation}

The values of \dndetanormknee and \sigmanormknee are fixed in Eq.~\ref{eq:ffunction} using the values reported in Table~\ref{tab:knee_sigma0}. Consequently, the function presented in Eq.~\ref{eq:ffunction} depends on \dndetanorm. The $f(\dndetanorm)$ has a rather simple behavior: $f(\dndetanorm) = \dndetanorm$ in the limit when \dndetanorm is smaller than the ratio of particle densities at the knee. Thus, Eq.~\ref{eq:fitmeanpt} becomes, $\ptnorm =1$ in this limit. Conversely,  when \dndetanorm is larger than the ratio of particle densities at the knee, $f(\dndetanorm)\approx \dndetanormknee$, and Eq.~\ref{eq:fitmeanpt} becomes: $\ptnorm \propto ( \dndetanorm / \dndetanormknee)^{\scs}$. For example, for the \VZERO-based centrality estimator, $f(\dndetanorm)$ is equal to 1.099 at the knee.

Table~\ref{tab:knee_sigma0} presents the obtained \scs values, along with their statistical and total systematic uncertainties, for each centrality estimator. The statistical uncertainty is calculated by independently shifting each \ptnorm data point by a fraction of its statistical uncertainty. Each fraction is randomly drawn from a standard normal distribution, and each new \ptnorm versus \dndetanorm correlations is refitted. This procedure is repeated a thousand times, resulting in a distribution of \scs values. The \scs distribution is then fitted with a Gaussian function, and its variance is associated with the statistical uncertainty on \scs. Two sources of systematic uncertainty are considered. The first source arises from the choice of the Boltzmann-Gibbs Blast-Wave~\cite{Schnedermann:1993ws} model to extrapolate the \pt spectra. To assess this, the \scs is extracted using alternative models: the Lévy-Tsallis~\cite{Wilk:1999dr} and Hagedorn~\cite{Hagedorn:1983wk} (described in Sec.~\ref{sec:systematic_uncertainties}). The maximum difference between \scs values obtained using the nominal and alternative models is assigned as the systematic uncertainty. The second source stems from the imprecise measurement of \sigmanormknee and \dndetanormknee. This is assessed by incoherently shifting these parameters around their nominal values. The shift amount is determined by their uncertainty multiplied by a random factor drawn from a Gaussian distribution with a mean of zero and standard deviation of one. Then, the \ptnorm distribution is fit for each shift. This process is repeated a thousand times, generating a distribution of \scs values. The standard deviation of this distribution, fitted with a Gaussian function, is assigned as the systematic uncertainty due to the imprecise measurement of knee parameters. The total systematic uncertainty on the value of \scs is obtained by summing in quadrature the systematic uncertainties from model dependence and knee parameter uncertainty.

\begin{figure}[!ht]
	\centering
	\hspace{0cm}
    \includegraphics[width=1\textwidth]{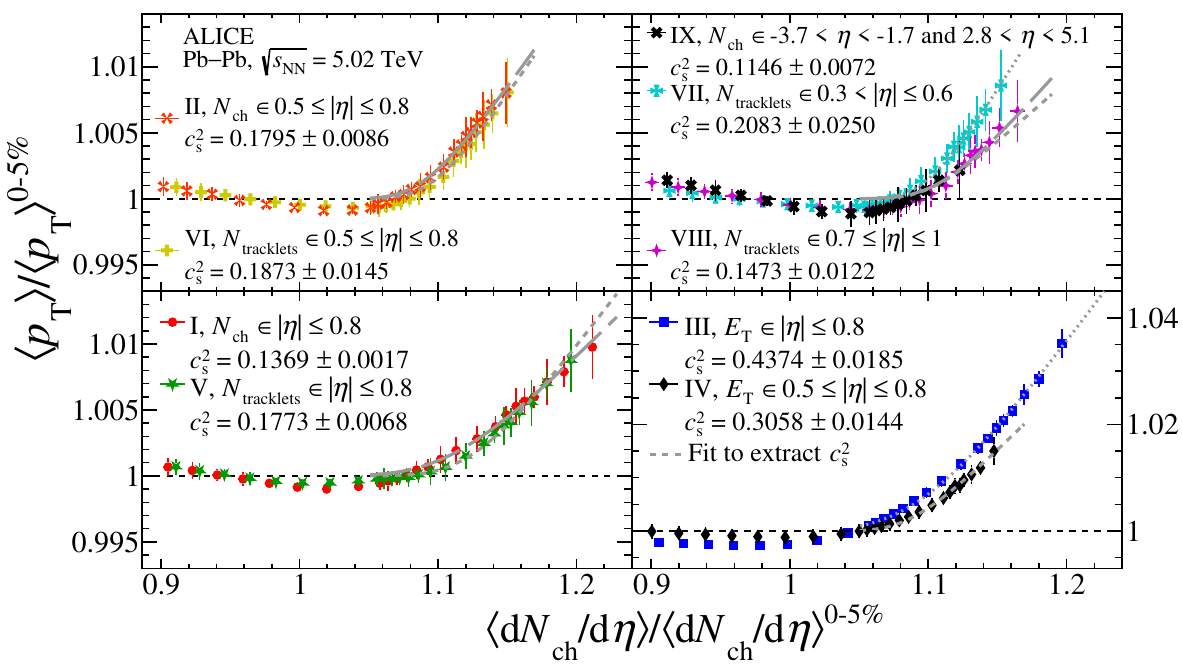}
	\hspace{0cm}
 	\caption{Correlation between \ptnorm and \dndetanorm in \PbPb collisions at \fivenn. Each panel shows the results for different centrality estimators defined in Table~\ref{tab:centrality_definition}. \ptnorm and \dndetanorm are derived from the \pt spectra within the 0--10 $\mathrm{GeV}/c$ interval for all centrality estimators. For \et-based centrality estimators, the $y$-axis scale should be read from the axis located to the right of the bottom right panel. Lines on top of the data represent fits using Eq.~\ref{eq:fitmeanpt}. The fit range spans from 1 to the last point. Each panel displays the corresponding \scs values with their total uncertainty, determined by summing the statistical and systematic uncertainties in quadrature. Vertical uncertainty bars in each point represent the combined statistical and systematic uncertainty. The total uncertainty in the \dndetanorm is negligible and therefore not visible.}
	\label{fig:meanpT_with_fits}
\end{figure}

Figure~\ref{fig:meanpT_with_fits} shows the \ptnorm as a function of \dndetanorm for the different centrality estimators with fits to the data using Eq.~\ref{eq:fitmeanpt}. The top left panel shows the results with pseudorapidity gap: centrality estimated in $0.5 \leq |\eta| < 0.8$, and \meanpt and \avdndeta measured in $|\eta|\leq0.3$. The event activity is quantified using \nch and \ntracklets in the \TPC and \SPD detectors, respectively. Both centrality estimators give similar results suggesting that the yield of particles with transverse momentum below $\pt=0.15~\mathrm{GeV}/c$ has a negligible impact on selecting collisions with similar entropy densities when using these two centrality estimators. This is further supported by the similar \scs values obtained with both estimators: \scs is equal to $0.1873 \pm 0.0145$ and $0.1795\pm 0.0086$ for the \SPD and \TPC centrality estimators, respectively. 

The top right panel of Fig.~\ref{fig:meanpT_with_fits} presents the results obtained with varying pseudorapidity gaps (0, 0.4, and 0.9 units). \ptnorm with zero gap rises at a steeper rate compared to the cases with a gap, while \dndetanorm remains relatively constant regardless of the presence of the pseudorapidity gap. This can be attributed to the finite width of jets, whose fragmentation products can extend into the pseudorapidity region where \meanpt is measured. The rise rate of \ptnorm is seen to decrease with increasing pseudorapidity gap, yielding \scs values of $0.2083\pm0.0250$ (no gap), $0.1873\pm0.0145$ (gap of 0.2 units) and $0.1473\pm0.0122$ (gap of 0.4 units), for the three cases using the \ntracklets centrality estimator. While it would be interesting to investigate larger pseudorapidity gaps with the \SPD detector, the strong dependence of cluster reconstruction efficiency on the primary-vertex position along the beam axis limits the maximum achievable pseudorapidity selection to one unit. The maximum pseudorapidity gap is achieved using the \VZERO detector. The \VZERO amplitude-based centrality estimator exhibits a narrower range in \dndetanorm, reaching only up to 1.1, compared to the particle density obtained with \SPD and \TPC centrality estimators. It is known that the \VZERO detector has a better centrality resolution than central-barrel detectors~\cite{ALICE:2013hur}. \ptnorm with the \VZERO centrality estimator also increases, albeit at a lower rate, yielding the lowest speed of sound, $\scs = 0.1146\pm0.0072$. Accordingly, the data suggest a decreasing trend in the extracted \scs with increasing pseudorapidity gap width. 

The bottom left panel of Fig.~\ref{fig:meanpT_with_fits} shows \ptnorm using multiplicity-based centrality estimators (\nch in the \TPC and \ntracklets in the \SPD). In this case, there is no pseudorapidity gap, and the pseudorapidity regions for centrality estimation and transverse-momentum spectra measurement fully overlap $(|\eta|\leq 0.8)$. The \dndetanorm reach with the \nch estimator is the highest among all centrality estimators. This is attributed to the fact that the same particles are used for both centrality determination and \pt spectra measurement within the same pseudorapidity region. Using overlapping pseudorapidity intervals introduces a multiplicity bias. Local fluctuations combined with this bias, including measurement uncertainties, lead to a broader  distribution along the \dndetanorm axis, while \ptnorm remains relatively unchanged, resulting in a lower extracted \scs ($0.1369\pm0.0017$ compared to $0.1795 \pm 0.0086$ with a pseudorapidity gap of 0.2 units). The \ptnorm distribution obtained using the \ntracklets-based centrality estimator with full overlap also appears to be stretched along the \dndetanorm axis compared to the cases with pseudorapidity gap. This results in a steeper \ptnorm and a larger extracted \scs compared to the \nch estimator. These results can be partially explained by the overlap between \SPD tracklets and global tracks, causing a significant multiplicity bias but  still smaller than the one introduced by estimating both centrality and \dndetanorm using the same pool of global tracks.

The bottom right panel of Fig.~\ref{fig:meanpT_with_fits} shows the results obtained using the transverse energy-based centrality estimators. The centrality estimator with full pseudorapidity overlap $(|\eta|\leq0.8)$ introduces a transverse-momentum bias leading to the largest \ptnorm. Furthermore, the \dndetanorm reaches 1.18, suggesting an additional multiplicity bias due to jet fragmentation. Introducing a 0.2 unit pseudorapidity gap reduces \dndetanorm to around 1.14, consistent with the values obtained using \SPD and \TPC multiplicity-based estimators with the same gap. Moreover, the \ptnorm rise becomes less steep compared to the case with full overlap, although the extracted \scs is higher than that obtained with multiplicity-based estimators. This could be attributed to an interplay between the finite width of  jets and the transverse-momentum bias. Finally, the different \ptnorm distributions measured using the multiplicity-based centrality estimators exhibit a shallow local minimum at $\dndetanorm \approx 1.04$. This feature is less noticeable when the \et-based centrality estimator is employed. The minimum corresponds to collisions in the 0.9--1\% centrality range. In Ref.~\cite{Nijs:2023bzv}, it is discussed that while the average entropy density, average temperature at hydrodynamic initialization, QGP size, and impact parameter all vary monotonically near 1\% centrality, the impact parameter essentially stops changing below 1\% centrality. This observation is consistent with the observed centrality percentile at which the plateau in the \avNpart distribution commences (see Fig.~\ref{fig:Npart_vs_cent}). Therefore, the local minimum could be attributed to the impact parameter ceasing to vary, or varying only minimally.

Appendix~\ref{app:meanpT_high_pT_cut} shows the correlation between \ptnorm and \dndetanorm using the \et-based and \nch-based centrality estimators. In both cases, a minimum pseudorapidity gap of 0.2 units was maintained between the pseudorapidity region used for centrality estimation and the region used to measure \ptnorm and \dndetanorm. The \et-based centrality estimator uses charged particles with a lower \pt threshold of $0.15~\mathrm{GeV}/c$, while the \nch-based centrality estimator uses a lower \pt threshold of $0.45~\mathrm{GeV}/c$. This selection is motivated by a prediction from the Trajectum model~\cite{Nijs:2023yab,Nijs:2020roc,Nijs:2021clz}: defining an \nch-based centrality estimator using charged particles with a relatively high \pt threshold selects collisions with a higher \ptnorm, compared to using a lower \pt threshold. A fit to the \ptnorm versus \dndetanorm correlation with the \nch-based centrality estimator and with a minimum pseudorapidity gap of 0.2 units yields an extracted \scs value that is about 27\% higher when the lower \pt threshold is set to $0.45~\mathrm{GeV}/c$ compared to the value obtained using \ntracklets for centrality estimation with a lower \pt threshold of $0.03~\mathrm{GeV}/c$ and with the same pseudorapidity gap. Furthermore, it is shown that the extraction of the \scs is robust against variations in the fit range.

\begin{figure}[!ht]
	\centering
	\hspace{0cm}
    \includegraphics[width=1\textwidth]{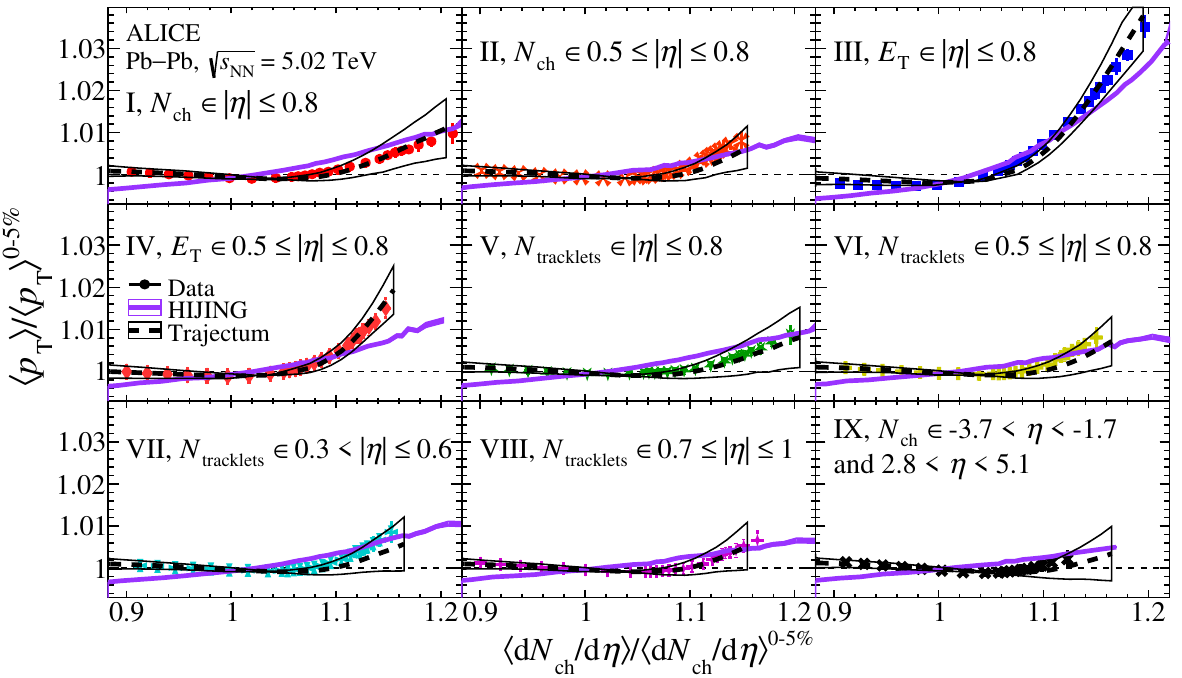}
	\hspace{0cm}
	\caption{Correlation between \ptnorm and \dndetanorm in \PbPb collisions at \fivenn. Each panel shows the results for different centrality estimators defined in Table~\ref{tab:centrality_definition}. The data are compared with predictions from the HIJING~\cite{PhysRevD.44.3501} and Trajectum~\cite{Nijs:2023yab,Nijs:2020roc,Nijs:2021clz} models, represented by continuous and dashed lines, respectively. The bands around the Trajectum predictions represent the sum in quadrature of the statistical and systematic uncertainties, with the latter being the dominant source. For the HIJING predictions, only the statistical uncertainty is shown (not visible in the plot).}
	\label{fig:MeanpT_vs_trajectum}
\end{figure}

Figure~\ref{fig:MeanpT_vs_trajectum} shows the \ptnorm versus \dndetanorm correlation in data compared to predictions by the HIJING~\cite{PhysRevD.44.3501} and Trajectum~\cite{Nijs:2023yab,Nijs:2020roc,Nijs:2021clz} models for the different centrality estimators defined in Table~\ref{tab:centrality_definition}. The HIJING model implements an independent source particle production picture, where each collision is modeled as a superposition of independent nucleon--nucleon collisions, neglecting interactions between the sources. Trajectum incorporates a generalized T$_\mathrm{\textnormal{R}}$ENTo~\cite{Moreland:2014oya} initial stage, followed by a viscous hydrodynamic stage and hadronization using SMASH~\cite{SMASH:2016zqf}. It utilizes a continuous hybrid equation of state (EoS) that interpolates between the hadron resonance gas (HRG) at low temperatures and the lattice result by HotQCD at high temperatures~\cite{HotQCD:2014kol}. Samples of ultracentral collisions are simulated using different EoS parameter settings~\cite{Nijs:2023bzv}. The dashed lines in Fig.~\ref{fig:MeanpT_vs_trajectum} represent the average Trajectum predictions, with the bands indicating the sum in quadrature of the statistical and systematic uncertainties. The latter are given by the one-standard-deviation uncertainty due to variations in EoS parameters. Trajectum predictions for \ptnorm versus \dndetanorm are in good agreement with the data, reproducing the minimum and the subsequent rise in \ptnorm and capturing the experimental biases of all centrality estimators, in particular the rapid rise of \ptnorm observed with the $E_{\mathrm{T}}$-based estimator. HIJING predicts a steady increase in \ptnorm with increasing \dndetanorm for all centrality estimators, with a steeper rise with the \et-based estimator with no pseudorapidity gap (less pronounced using the same estimator with a gap of 0.2 units). In the ultracentral-collision limit the \et-based estimator clearly favors events with an enhanced production of multiple jets~\cite{PhysRevD.44.3501}, leading to a larger \ptnorm. Since HIJING does not include QGP formation, its reproduction of the observed selection biases suggests that the rise of the \meanpt in ultracentral collisions may not directly probe entropy fluctuations in the QGP.

\begin{figure}[!ht]
	\centering
	\hspace{0cm}
    \includegraphics[width=1\textwidth]{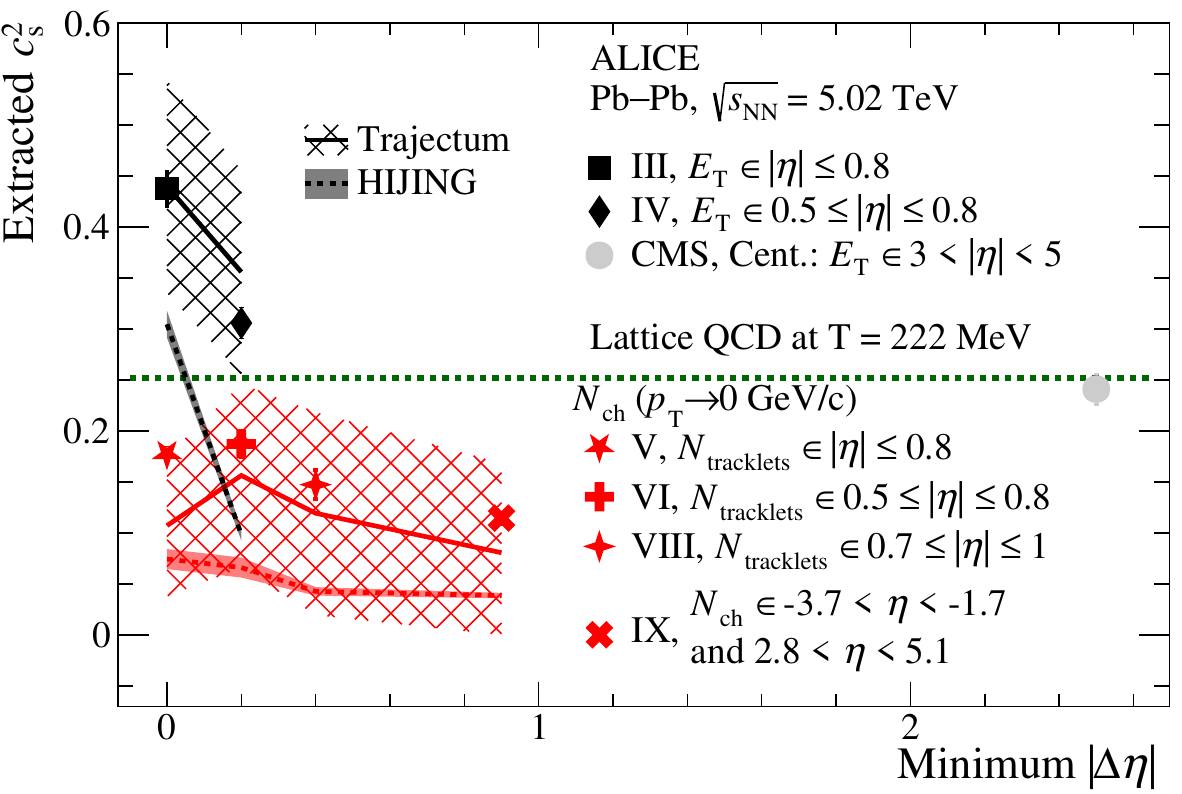}
	\hspace{0cm}
	\caption{Extracted \scs as a function of the minimum pseudorapidity gap between the centrality estimation and transverse-momentum spectra measurement in Pb--Pb collisions at \fivenn. The results are compared with the \scs value measured by the CMS Collaboration (with a minimum $|\Delta\eta|=2.5$)~\cite{CMS:2024sgx}. The uncertainty bars around the data points represent the sum in quadrature of the statistical and systematic uncertainties (not visible). Predictions from Trajectum~\cite{Nijs:2023yab,Nijs:2020roc,Nijs:2021clz} are shown as continuous lines, with mesh bands indicating the total uncertainty (statistical and systematic combined in quadrature). Predictions from the HIJING~\cite{PhysRevD.44.3501} model are shown as dashed lines, with shaded bands representing statistical uncertainty only. See the main text for details on the uncertainty estimation in the model \scs values. The horizontal dashed green line indicates the Lattice QCD prediction of \scs for deconfined matter, as calculated by the HotQCD Collaboration~\cite{HotQCD:2014kol}. Centrality-estimator definitions are presented in Table~\ref{tab:centrality_definition}.}
	\label{fig:cs2_vs_eta_gap}
\end{figure}

Figure~\ref{fig:cs2_vs_eta_gap} illustrates the dependence of the extracted \scs on the pseudorapidity gap between the centrality determination region and the transverse-momentum spectra measurement region. The highest \scs value is observed with the transverse energy-based centrality estimation and full pseudorapidity overlap. This configuration biases \ptnorm, potentially introducing a jet-fragmentation bias. Introducing a pseudorapidity gap in the transverse-energy centrality estimation reduces both \ptnorm and \dndetanorm, leading to a lower \scs of approximately 0.3. However, this value remains significantly higher than those obtained with multiplicity-based estimators, likely due to the contribution of intermediate-to-high \pt particles from jet fragmentation to the spectra. Notably, the multiplicity-based centrality estimator with the \SPD and full overlap yields a \scs $(0.1773 \pm 0.0068)$ similar to that obtained with a larger pseudorapidity gap $(0.1873 \pm 0.0145)$. This similarity arises from a multiplicity bias in the no-gap case, which stretches the \ptnorm versus \dndetanorm distribution along the \dndetanorm axis, resulting in a lower extracted \scs. When the centrality and the \pt spectrum are determined in non-overlapping regions, the \scs found with the \ntracklets centrality estimator decreases with increasing pseudorapidity gap. This dependence on the pseudorapidity gap is further supported by the even lower \scs values obtained with the \VZERO centrality estimator.

Figure~\ref{fig:cs2_vs_eta_gap} also presents the extracted \scs value by the CMS Collaboration~\cite{CMS:2024sgx}, where the centrality is determined in the forward rapidity region $(3\leq|\eta|\leq5)$ using the top 0--5\% most energetic events. The \ptnorm versus \dndetanorm correlation is measured in $|\eta|<0.5$. CMS reports a $\scs=0.241\pm 0.002~(\mathrm{stat.})~\pm~0.016~(\mathrm{syst.})$ in natural units, consistent with Lattice QCD expectations. This value falls between the \scs values obtained using the charged-particle multiplicity and transverse-energy centrality estimators in ALICE data. The CMS experimental setup employs a significantly wider pseudorapidity gap between the centrality and observable pseudorapidity regions compared to ALICE, effectively  suppressing short-range \meanpt--\meanpt correlations due to jet finite width. However, even with this gap, the \et-based centrality estimator remains sensitive to long-range \meanpt--\meanpt correlations~\cite{Chatterjee:2017mhc}.

Figure~\ref{fig:cs2_vs_eta_gap} shows the extracted \scs values using simulated events from the Trajectum~\cite{Nijs:2023yab,Nijs:2020roc,Nijs:2021clz} and HIJING~\cite{PhysRevD.44.3501} models. The model values are extracted by fitting the predicted \ptnorm versus \dndetanorm correlation using the same procedure and centrality estimators as for the data. The uncertainty in the Trajectum \scs values is estimated by shifting the \ptnorm distribution between its minimum and maximum bounds, as determined by its point-by-point systematic uncertainty, and then refitting. The difference between the minimum and maximum extracted \scs values, relative to the nominal value, is assigned as the systematic uncertainty. Only the statistical uncertainty is considered for the HIJING predictions. Trajectum predictions are consistent with the data within 15\% for the \et-based centrality estimator, but exhibit increasing deviation with increasing pseudorapidity gap for multiplicity-based centrality estimators. In contrast, the HIJING \scs values are systematically lower than the data, indicating that hydrodynamic evolution is essential for a quantitative description.

\begin{figure}[!ht]
	\centering
	\hspace{0cm}
    \includegraphics[width=0.95\textwidth]{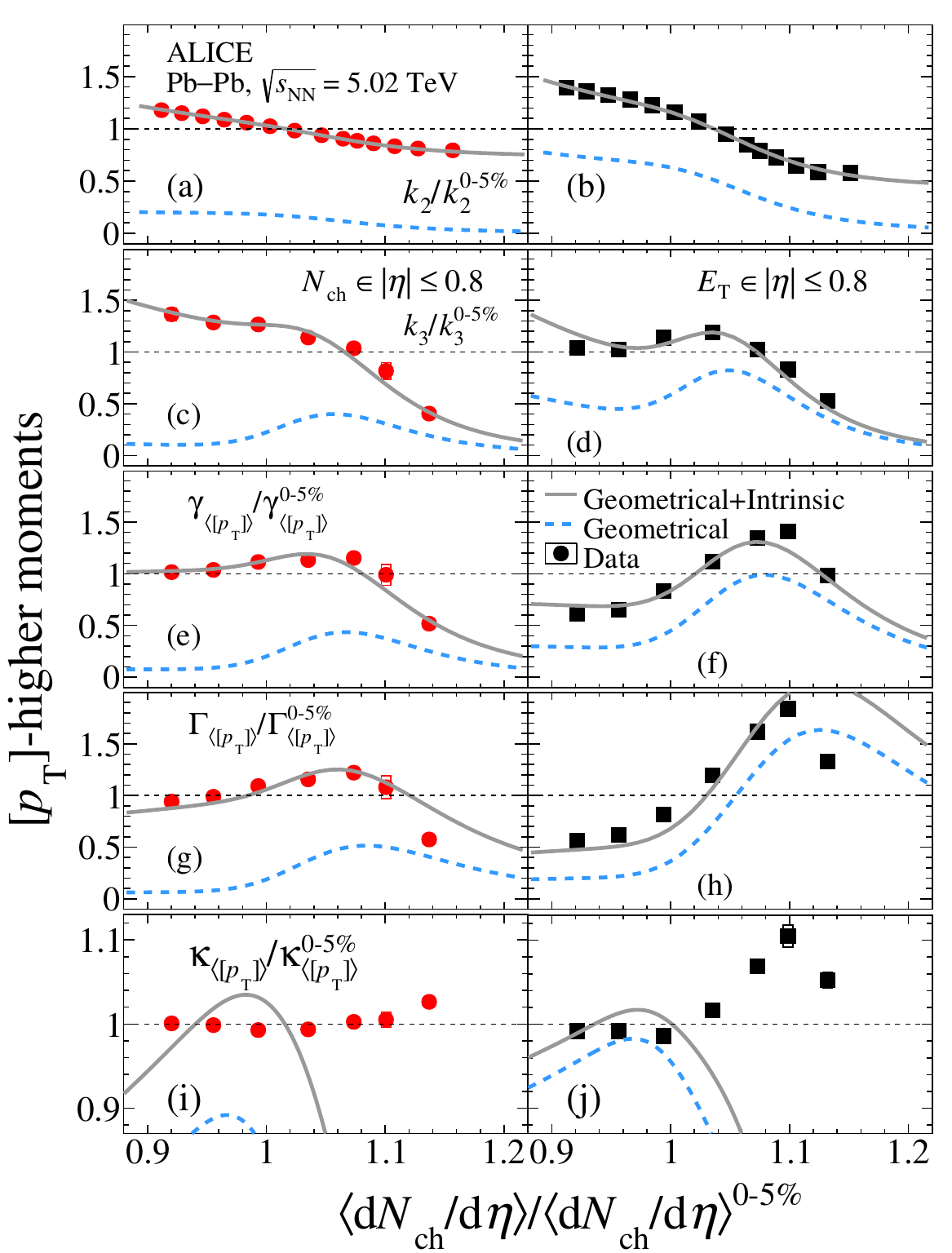}
	\hspace{0cm}
	\caption{Higher-order moments of $[\pt]$ measured for events selected with the \nch (I) and \et (III) centrality estimators at midrapidity in \PbPb collisions at \fivenn are shown in the left and right columns, respectively. The \ktwnorm distributions are fitted with a two-component model (continuous gray line), where the estimated Geometrical component is also shown (dotted-dashed blue line)~\cite{Samanta:2023amp}. The curves for the \kthnorm, \stdsknorm, \intsknorm, and \kunorm distributions correspond to predictions based on the fit of \ktwnorm~\cite{Samanta:2023kfk}.}
	\label{fig:higher_orders_fits}
\end{figure}

Figure~\ref{fig:higher_orders_fits} presents the self-normalized higher-order moments of $[\pt]$ as a function of \dndetanorm. The left column displays results from the charged-particle centrality estimator at midrapidity, while the right column shows those from the transverse-energy centrality estimator. Results using the V0M centrality estimator are similar to those from the midrapidity multiplicity-based centrality estimator and are shown in Appendix~\ref{app:highermoments_v0m}. 

Under the assumption of independent nucleon--nucleon particle production, higher-order moments of $[\pt]$ are expected to depend on collision centrality and system size, as characterized by \Npart. The $n^{\mathrm{th}}$-order cumulant is expected to scale as $\propto 1/\Npart^{(n-1)}$, or equivalently as $\propto 1/\nch^{(n-1)}$~\cite{Giacalone:2020lbm,Bhatta:2021qfk}. Consequently, $\ktwnorm \propto 1/\Npart$. The \ktwnorm distributions in panels (a) and (b) exhibit a decreasing trend with increasing \dndetanorm. While the \ktwnorm\,decreases proportionally to $1/\dndetanorm$ with the \nch-based centrality estimator, the decrease is more rapid with the \et-based estimator for $\dndetanorm \lesssim 1.1$. The non-trivial evolution of \ktwnorm with the \et estimator can be attributed to the interplay of two effects: jet-fragmentation bias, arising from the overlap of centrality assessment and \ktwnorm measurement within the same pseudorapidity window, and volume fluctuations, stemming from the centrality-dependent \avNpart (see Fig.~\ref{fig:Npart_vs_cent}). The \et estimator selects collisions with lower \avNpart compared to the \nch estimator, resulting in larger volume fluctuations.

For a fixed \avdndeta, the fluctuations of $[\pt]$, quantified by its variance, $Var($[\pt]$|\avdndeta)$, arise from both volume and quantum fluctuations. Volume fluctuations originate from impact parameter variations and are referred to as Geometrical fluctuations. Additionally, $[\pt]$ can fluctuate even when both \avdndeta and $b$ are constant, these are referred to as Intrinsic fluctuations. The total variance can be expressed as $Var([\pt]|\avdndeta) = ( \langle \overline{[\pt]}^{2} \rangle_b - \langle \overline{[\pt]}\rangle^{2}_b ) + \langle Var([\pt]|\avdndeta) \rangle_b$, where $\overline{[\pt]}$ represents the expected value of $[\pt]$ as a function of $b$ and \avdndeta, and $\langle \cdots \rangle$ denotes an average over $b$. The first term describes Geometrical fluctuations, while the second term represents the Intrinsic ones~\cite{Samanta:2023amp}.

To fit the \ktwnorm distribution using the two-component model described above, it is assumed that the joint probability of \dndetanorm and $[\pt]^{\mathrm{norm}}$ at fixed-impact parameter is given by a two-dimensional Gaussian~\cite{Samanta:2023amp} distribution. This distribution is defined by five parameters: the mean and variance of  $[\pt]^{\mathrm{norm}}$ and \dndetanorm, denoted by $\overline{[\pt]^{\mathrm{norm}}}(b)$, $\overline{\dndetanorm}(b)$, $Var([\pt]^{\mathrm{norm}}|b)$, $Var(\dndetanorm|b)$, and the correlation coefficient, $r(b)$, between $[\pt]^{\mathrm{norm}}$ and \dndetanorm. The $\overline{\dndetanorm}(b)$ and $Var(\dndetanorm|b)$, or equivalently, $\sigma^{\mathrm{norm}}(b)$, are obtained from fitting the event fraction distribution as a function of \dndetanorm, as described earlier (see Fig.~\ref{fig:event_fraction}). Since the event-average transverse momentum is independent of collision centrality for the 30\% most central collisions~\cite{ALICE:2018hza}, $\overline{[\pt]^{\mathrm{norm}}}(b)$ is assumed to be independent of $b$ and is denoted by $\overline{[\pt]^{\mathrm{norm}}}_{0}$. The variance of $[\pt]^{\mathrm{norm}}$ for fixed $b$ is given by $Var([\pt]^{\mathrm{norm}}|b)= (1 - r(b)^{2}) \sigma_{[\pt]}^{2}(\dndetanormknee / \overline{\dndetanorm}(b))^{\alpha}$, where $\sigma_{[\pt]}$ is the standard deviation of $[\pt]^{\mathrm{norm}}$ for collisions at zero-impact parameter, and $\alpha$ describes the decrease of the variance as a function of impact parameter. The parameters $\sigma_{[\pt]}, \alpha$ , and the correlation coefficient are treated as fit parameters and are assumed to be independent of the impact parameter.

Fits to the \ktwnorm distributions are performed separately for the \nch- and \et-based centrality estimators. The obtained values with the \nch estimator are: $\sigma_{[\pt]}=4.780, \alpha = 1.613$, and $ r=0.921$, while for the \et-based centrality estimator they are: $\sigma_{[\pt]}=14.539, \alpha = 2.512$, and $ r=0.985$. The most significant difference between the two sets of parameters is observed for $\sigma_{[\pt]}$, which is about three times larger for the \et-based centrality estimator than for the \nch-based estimator. Fits to the data suggest that the \et-based centrality estimator preferentially selects events where $[\pt]$ fluctuations primarily originate from Geometrical fluctuations. The relative contribution of the Geometrical component to $[\pt]$ fluctuations decreases from 18\% (55\%) at $\dndetanorm\approx1$ to approximately 6\% (24\%) at $\dndetanorm\approx1.14$ for the \nch-(\et-)based centrality estimators. In contrast, the model suggests that $[\pt]$ variations primarily originate from Intrinsic fluctuations when using the \nch-based centrality estimator. However, in the ultracentral-collision limit, the variance decreases dramatically, which can be explained by a significant suppression of the Geometrical component in this regime.

The ATLAS Collaboration has reported higher-order moments of $[\pt]$ in ultracentral collisions selected using the forward $\Sigma \et$~\cite{ATLAS:2024jvf}. The ATLAS results exhibit closer qualitative agreement with those obtained in this study for the \et-based centrality estimator, further supporting the dominance of the Geometrical component. 

Panels (c) to (h) of Fig.~\ref{fig:higher_orders_fits} show the \kthnorm ((c) and (d)), \stdsknorm ((e) and (f)), and \intsknorm ((g) and (h)). Under the assumption of independent nucleon--nucleon particle production, \kthnorm is expected to scale as $\kthnorm \propto 1/\Npart^{2}$ and $\stdsknorm \propto 1/\sqrt{\Npart}$, while the centrality dependence of \intsknorm is expected to be milder than for \stdsknorm~\cite{Giacalone:2020lbm}. 

Skewness $(\kthnorm)$ and kurtosis encode the non-Gaussian properties of the event-by-event $[\pt]$ distribution. With the \nch-based centrality estimator, \kthnorm decreases with increasing centrality, while a slight increase is observed around $\dndetanorm\approx1.04$. The standardized $(\stdsknorm)$ and intensive $(\intsknorm)$ skewness distributions exhibit an increase with \dndetanorm, reaching a maximum around the knee region, followed by a decreasing trend towards the ultracentral-collision limit. The results with the \et-based centrality estimator exhibit similar features, but with more pronounced maxima. The observed \dndetanorm dependence deviates from the expectations under the assumption of independent particle sources. The model predictions, overlaid on the distributions, demonstrate a dependence on the centrality estimator. The model accurately predicts the observed increase and the maximum in the standardized and intensive skewness distributions, except for the intensive skewness with the \et-based centrality estimator. 

Panels (i) and (j) of Fig.~\ref{fig:higher_orders_fits} show the kurtosis $(\kunorm)$ of the event-by-event $[\pt]$ distribution for \nch and \et centrality estimators, respectively. The kurtosis decreases slightly below $\dndetanorm=1$ before increasing towards the ultracentral collisions. While the charged-particle multiplicity results show only a modest increase in the ultracentral-collision limit, the \et-based results exhibit a peak around $\dndetanorm \approx 1.1$. The predicted \kunorm for both centrality estimators show a dependence on \dndetanorm peaking below \dndetanormknee, with a shape very different from the relatively flat evolution of the data for the entire \dndetanorm.

A key limitation of the model lies in its assumption of a Gaussian distribution of $[\pt]$ at fixed-impact parameter. Since $[\pt]$ is a positive quantity, its distribution is expected to exhibit positive skewness and kurtosis. This inherent property of the $[\pt]$ distribution will contribute to the observed skewness and kurtosis~\cite{Samanta:2023kfk}.

\begin{figure}[!ht]
	\centering
	\hspace{0cm}
    \includegraphics[width=0.95\textwidth]{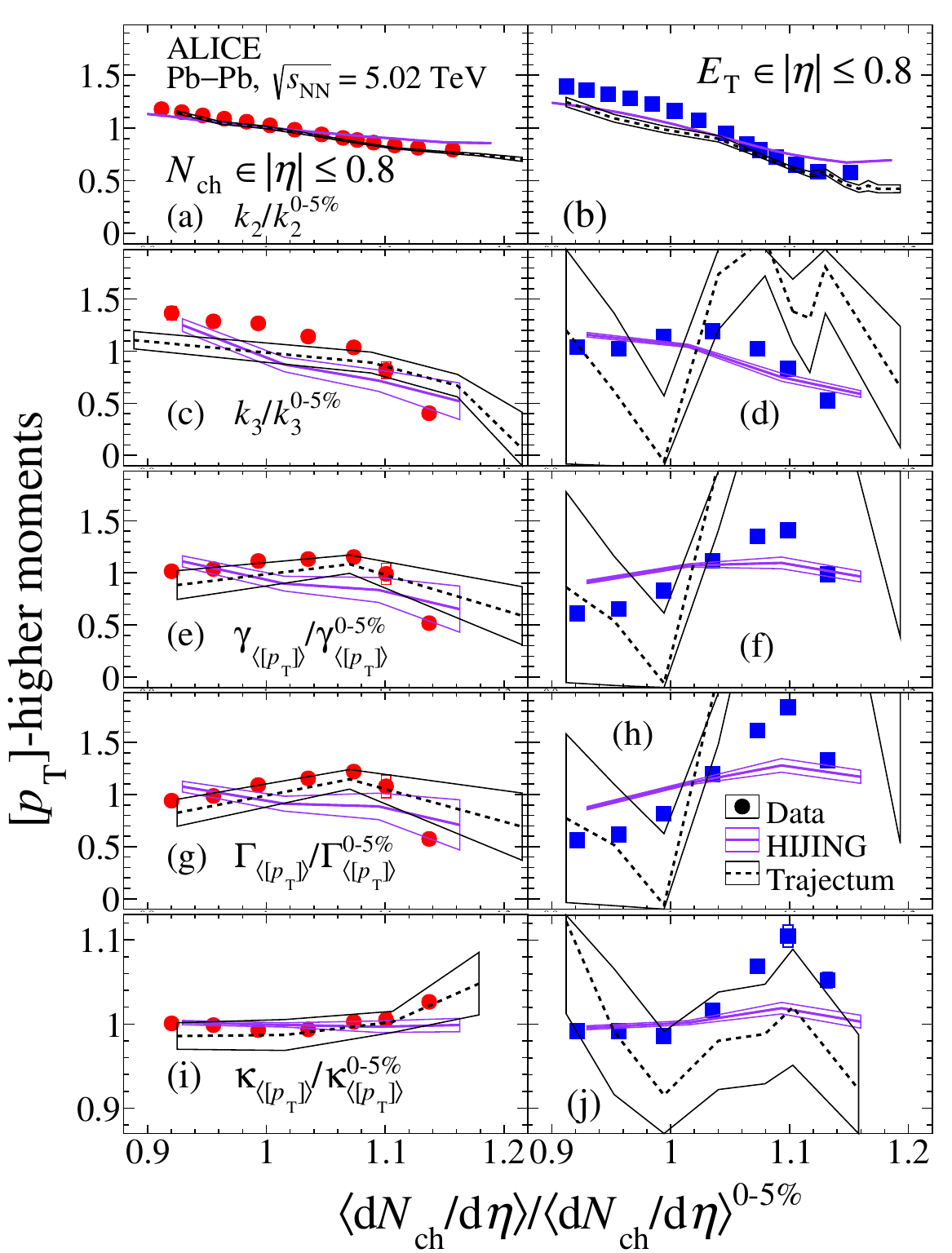}
	\hspace{0cm}
	\caption{Higher-order moments of $[\pt]$ for events selected with the \nch (I) and \et (III) centrality estimators at midrapidity in \PbPb collisions at \fivenn are shown in the left and right columns, respectively. The data are compared with predictions from the HIJING~\cite{PhysRevD.44.3501} and Trajectum~\cite{Nijs:2023yab,Nijs:2020roc,Nijs:2021clz} models. The width of the bands represents the statistical uncertainty in the HIJING predictions. The bands around the Trajectum predictions represent the sum in quadrature of the statistical and systematic uncertainties, with the latter being the dominant source.}
	\label{fig:higher_orders_models}
\end{figure}

Figure~\ref{fig:higher_orders_models} compares the self-normalized higher-order moments of $[\pt]$  with predictions from the HIJING~\cite{PhysRevD.44.3501} and Trajectum~\cite{Nijs:2023yab,Nijs:2020roc,Nijs:2021clz} model. Predictions are shown for the \nch- and \et-based centrality estimators at midrapidity. Within the independent source picture implemented in the HIJING model, the $n^{\mathrm{th}}$-order cumulant is expected to scale as $\propto 1/\nch^{(n-1)}$~\cite{Bhatta:2021qfk}. However, this approach fails to accurately describe the event-to-event fluctuations of the average transverse momentum observed across the range from peripheral to central collisions~\cite{ALICE:2014gvd}. The shape of the predicted \ktwnorm for both centrality estimators qualitatively agrees with the data. Notably, HIJING predicts a more rapid decrease of \ktwnorm for $\dndetanorm \gtrsim 1$ with the \et-based centrality estimator, similar to what is seen in data. However, the model fails to reproduce the observed bumpy structure in the \kthnorm, \stdsknorm,  and \intsknorm distributions around $\dndetanorm \approx 1.08$. Finally, HIJING agrees with the kurtosis data below $\dndetanorm=1$ but underestimates the rise 
in the ultracentral collision regime. The hydrodynamic Trajectum model accurately describes the higher-order $[\pt]$ fluctuations for the \nch-based centrality estimator and captures the decrease in $\ktwnorm$, the peaks in $\stdskewnorm$ and $\intskewnorm$, and the rise in $\kurnorm$ in ultra-central collisions. The \et-based predictions rely on a considerably smaller sample than the \nch-based ones. Consequently, the substantial uncertainties in the Trajectum model's predictions for skewness and kurtosis preclude drawing conclusions. Nevertheless, the rapid decrease of \ktwnorm is well described in the ultracentral-collision limit.


\section{Conclusions}
\label{sec:Conclusions}
This study investigates the dependence of \ptnorm on \dndetanorm measured at midrapidity in ultracentral Pb--Pb collisions at \fivenn using different centrality estimators based on \nch and \et. Our findings reveal that the \et-based centrality estimator leads to a steeper and higher \ptnorm, potentially influenced by jet-fragmentation biases. Utilizing the \nch-based centrality estimator mitigates these biases, and introducing a pseudorapidity gap between the centrality and \ptnorm measurement regions further reduces their impact. The extracted $c_{\mathrm{s}}^{2}$ is found to strongly depend on the exploited centrality estimator and ranges between $0.1146 \pm 0.0028 \, \mathrm{(stat.)} \pm 0.0065 \, \mathrm{(syst.)}$ and $0.4374 \pm 0.0006 \, \mathrm{(stat.)} \pm 0.0184 \, \mathrm{(syst.)}$ in natural units. Based on HIJING-model predictions, which show a steady rise of \meanpt with $\langle \nch \rangle$, the observed increase of \ptnorm in the ultracentral-collision limit cannot be solely attributed to fluctuations in the initial state. Consequently, the extracted \scs may not directly correspond to the speed of sound in the QGP. These measurements confirm a prediction from the Trajectum hydrodynamic model~\cite{Nijs:2023bzv}, and necessitate a reevaluation of how the speed of sound can be extracted from heavy-ion data.

This study also presents measurements of higher-order moments of $[\pt]$ in ultracentral collisions, focusing on comparisons between \nch and \et centrality estimators at midrapidity. \kthnorm, \stdsknorm, and \intsknorm exhibit an increase around \dndetanormknee followed by a decrease towards the ultracentral-collision limit. These observations deviate from expectations based on independent particle production sources, as implemented in the HIJING model. The two-component model combining Intrinsic and Geometric sources of $[\pt]$ fluctuations~\cite{Samanta:2023amp,Samanta:2023kfk} reproduces most features of the data and suggests that the observed maxima in skewness variables may be attributed to Geometrical fluctuations in the initial state, which vanish in the ultracentral-collision limit.


\newenvironment{acknowledgement}{\relax}{\relax}
\begin{acknowledgement}
\section*{Acknowledgements}

The ALICE Collaboration would like to thank all its engineers and technicians for their invaluable contributions to the construction of the experiment and the CERN accelerator teams for the outstanding performance of the LHC complex.
The ALICE Collaboration gratefully acknowledges the resources and support provided by all Grid centres and the Worldwide LHC Computing Grid (WLCG) collaboration.
The ALICE Collaboration acknowledges the following funding agencies for their support in building and running the ALICE detector:
A. I. Alikhanyan National Science Laboratory (Yerevan Physics Institute) Foundation (ANSL), State Committee of Science and World Federation of Scientists (WFS), Armenia;
Austrian Academy of Sciences, Austrian Science Fund (FWF): [M 2467-N36] and Nationalstiftung f\"{u}r Forschung, Technologie und Entwicklung, Austria;
Ministry of Communications and High Technologies, National Nuclear Research Center, Azerbaijan;
Rede Nacional de Física de Altas Energias (Renafae), Financiadora de Estudos e Projetos (Finep), Funda\c{c}\~{a}o de Amparo \`{a} Pesquisa do Estado de S\~{a}o Paulo (FAPESP) and The Sao Paulo Research Foundation  (FAPESP), Brazil;
Bulgarian Ministry of Education and Science, within the National Roadmap for Research Infrastructures 2020-2027 (object CERN), Bulgaria;
Ministry of Education of China (MOEC) , Ministry of Science \& Technology of China (MSTC) and National Natural Science Foundation of China (NSFC), China;
Ministry of Science and Education and Croatian Science Foundation, Croatia;
Centro de Aplicaciones Tecnol\'{o}gicas y Desarrollo Nuclear (CEADEN), Cubaenerg\'{\i}a, Cuba;
Ministry of Education, Youth and Sports of the Czech Republic, Czech Republic;
The Danish Council for Independent Research | Natural Sciences, the VILLUM FONDEN and Danish National Research Foundation (DNRF), Denmark;
Helsinki Institute of Physics (HIP), Finland;
Commissariat \`{a} l'Energie Atomique (CEA) and Institut National de Physique Nucl\'{e}aire et de Physique des Particules (IN2P3) and Centre National de la Recherche Scientifique (CNRS), France;
Bundesministerium f\"{u}r Bildung und Forschung (BMBF) and GSI Helmholtzzentrum f\"{u}r Schwerionenforschung GmbH, Germany;
General Secretariat for Research and Technology, Ministry of Education, Research and Religions, Greece;
National Research, Development and Innovation Office, Hungary;
Department of Atomic Energy Government of India (DAE), Department of Science and Technology, Government of India (DST), University Grants Commission, Government of India (UGC) and Council of Scientific and Industrial Research (CSIR), India;
National Research and Innovation Agency - BRIN, Indonesia;
Istituto Nazionale di Fisica Nucleare (INFN), Italy;
Japanese Ministry of Education, Culture, Sports, Science and Technology (MEXT) and Japan Society for the Promotion of Science (JSPS) KAKENHI, Japan;
Consejo Nacional de Ciencia (CONACYT) y Tecnolog\'{i}a, through Fondo de Cooperaci\'{o}n Internacional en Ciencia y Tecnolog\'{i}a (FONCICYT) and Direcci\'{o}n General de Asuntos del Personal Academico (DGAPA), Mexico;
Nederlandse Organisatie voor Wetenschappelijk Onderzoek (NWO), Netherlands;
The Research Council of Norway, Norway;
Pontificia Universidad Cat\'{o}lica del Per\'{u}, Peru;
Ministry of Science and Higher Education, National Science Centre and WUT ID-UB, Poland;
National Research Foundation of Korea (NRF), Republic of Korea;
Ministry of Education and Scientific Research, Institute of Atomic Physics, Ministry of Research and Innovation and Institute of Atomic Physics and Universitatea Nationala de Stiinta si Tehnologie Politehnica Bucuresti, Romania;
Ministerstvo skolstva, vyskumu, vyvoja a mladeze SR, Slovakia;
National Research Foundation of South Africa, South Africa;
Swedish Research Council (VR) and Knut \& Alice Wallenberg Foundation (KAW), Sweden;
European Organization for Nuclear Research, Switzerland;
Suranaree University of Technology (SUT), National Science and Technology Development Agency (NSTDA) and National Science, Research and Innovation Fund (NSRF via PMU-B B05F650021), Thailand;
Turkish Energy, Nuclear and Mineral Research Agency (TENMAK), Turkey;
National Academy of  Sciences of Ukraine, Ukraine;
Science and Technology Facilities Council (STFC), United Kingdom;
National Science Foundation of the United States of America (NSF) and United States Department of Energy, Office of Nuclear Physics (DOE NP), United States of America.
In addition, individual groups or members have received support from:
Czech Science Foundation (grant no. 23-07499S), Czech Republic;
FORTE project, reg.\ no.\ CZ.02.01.01/00/22\_008/0004632, Czech Republic, co-funded by the European Union, Czech Republic;
European Research Council (grant no. 950692), European Union;
Deutsche Forschungs Gemeinschaft (DFG, German Research Foundation) ``Neutrinos and Dark Matter in Astro- and Particle Physics'' (grant no. SFB 1258), Germany;
ICSC - National Research Center for High Performance Computing, Big Data and Quantum Computing and FAIR - Future Artificial Intelligence Research, funded by the NextGenerationEU program (Italy).

\end{acknowledgement}

\bibliographystyle{utphys}   
\bibliography{bibliography}

\newpage
\appendix

\section{\texorpdfstring{$\mathbf{\langle \textit{p}_{\mathrm{T}}\rangle}/\mathbf{\langle \textit{p}_{\mathrm{T}}\rangle}^{0-5\%}$}{pT/pT(0-5)} as a function of \texorpdfstring{$\mathbf{\langle d\textit{N}_{ch}/d\eta \rangle}/\mathbf{\langle d\textit{N}_{ch}/d\eta \rangle}^{0-5\%}$}{dN/deta} using \texorpdfstring{$\mathbf{\textit{p}_{T}}$}{pT} selection} 
\label{app:meanpT_high_pT_cut}

\begin{figure}[!ht]
	\centering
	\hspace{0cm}
    \includegraphics[width=1\textwidth]{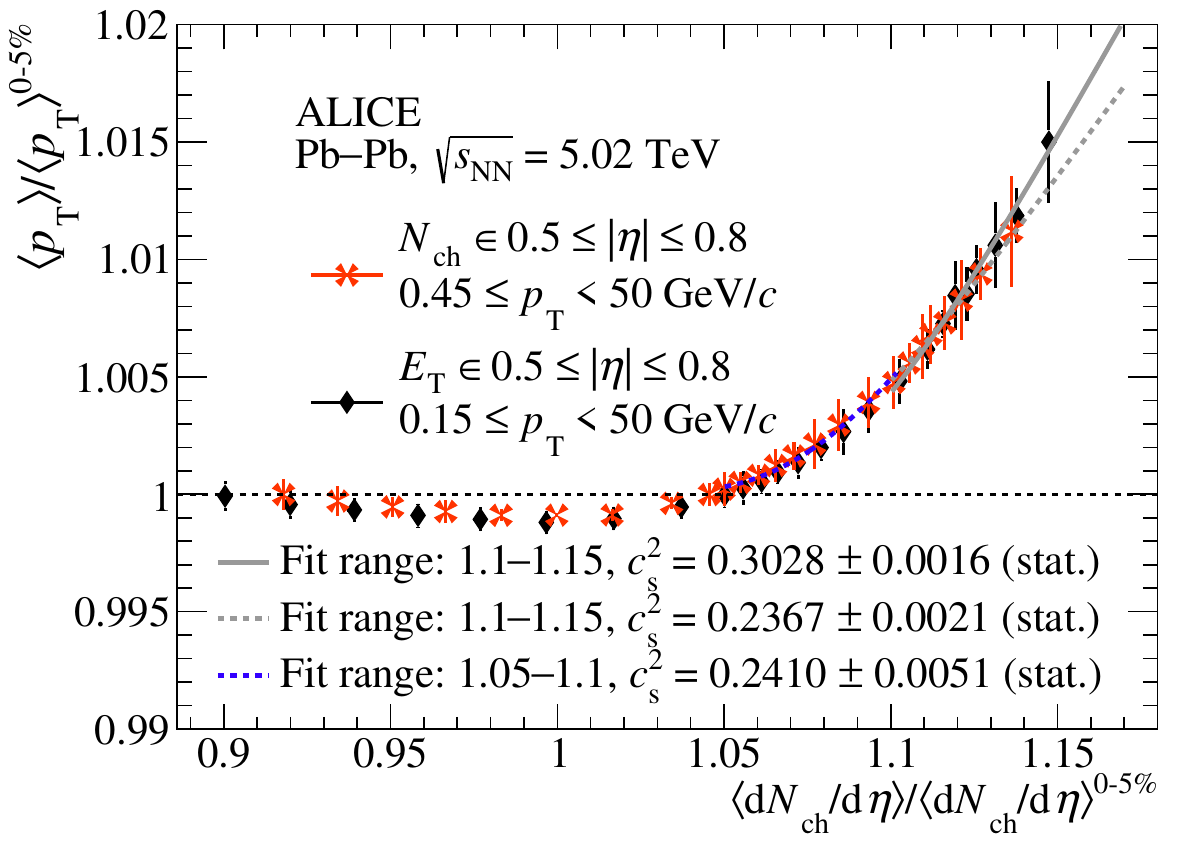}
	\hspace{0cm}
	\caption{\ptnorm as a function of \dndetanorm. Results are shown for two centrality estimators: one based on transverse energy (black diamonds) and the other based on charged-particle multiplicity (red crosses). For the \nch-based centrality estimator, two values of \scs are extracted using non-overlapping fit ranges (dashed-lines). The \ptnorm versus \dndetanorm correlation for the \et-based centrality estimator corresponds to the centrality definition labeled as IV in Table~\ref{tab:centrality_definition}.}
	\label{fig:meanpT_high_pT_cut}
\end{figure}

\newpage
\section{Higher-order moments of \texorpdfstring{$\mathbf{[\textit{p}_{T}]}$}{pT} moments with the V0M centrality estimator}
\label{app:highermoments_v0m}

\begin{figure}[!ht]
	\centering
	\hspace{0cm}
    \includegraphics[width=1\textwidth]{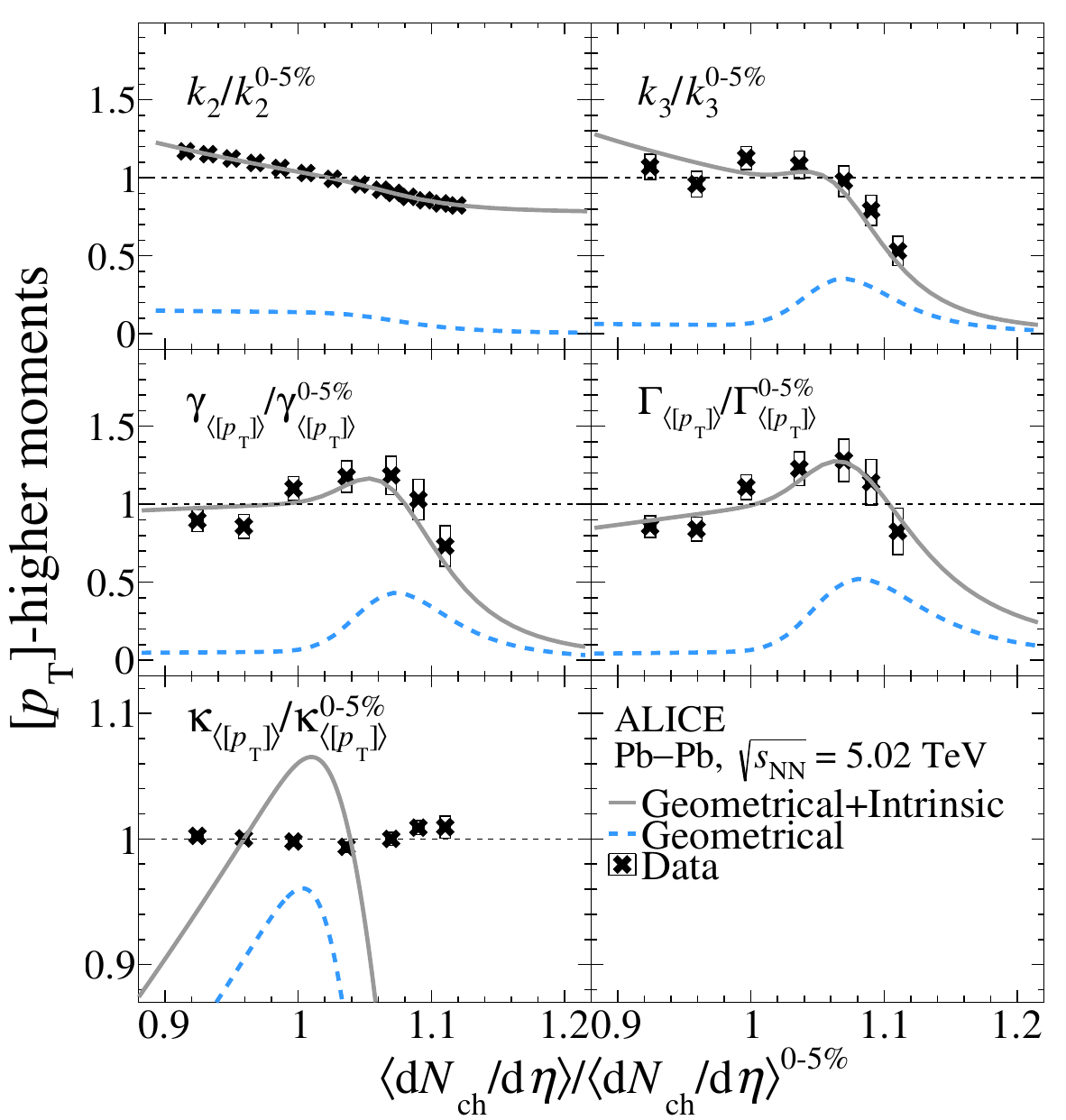}
	\hspace{0cm}
	\caption{Higher-order moments of $[\pt]$ measured for events selected with the \VZERO centrality estimator (IX) in \PbPb collisions at \fivenn. The \ktwnorm distribution is fitted with a two-component model (continuous gray line), where the estimated Geometrical component is also show (dotted-dashed blue line)~\cite{Samanta:2023amp}. The curves for the \kthnorm, \stdsknorm, \intsknorm, and \kunorm distributions correspond to predictions based on the fit of \ktwnorm~\cite{Samanta:2023kfk}.}
	\label{fig:higher_orders_fits_v0m}
\end{figure}

\newpage
\section{The ALICE Collaboration}
\label{app:collab}
\begin{flushleft} 
\small

I.J.~Abualrob\,\orcidlink{0009-0005-3519-5631}\,$^{\rm 113}$, 
S.~Acharya\,\orcidlink{0000-0002-9213-5329}\,$^{\rm 50}$, 
G.~Aglieri Rinella\,\orcidlink{0000-0002-9611-3696}\,$^{\rm 32}$, 
L.~Aglietta\,\orcidlink{0009-0003-0763-6802}\,$^{\rm 24}$, 
M.~Agnello\,\orcidlink{0000-0002-0760-5075}\,$^{\rm 29}$, 
N.~Agrawal\,\orcidlink{0000-0003-0348-9836}\,$^{\rm 25}$, 
Z.~Ahammed\,\orcidlink{0000-0001-5241-7412}\,$^{\rm 132}$, 
S.~Ahmad\,\orcidlink{0000-0003-0497-5705}\,$^{\rm 15}$, 
I.~Ahuja\,\orcidlink{0000-0002-4417-1392}\,$^{\rm 36}$, 
Z.~Akbar$^{\rm 80}$
% ...

\section*{Affiliation Notes}

$^{\rm I}$ Deceased\\
$^{\rm II}$ Also at: Max-Planck-Institut fur Physik, Munich, Germany\\
$^{\rm III}$ Also at: Italian National Agency for New Technologies, Energy and Sustainable Economic Development (ENEA), Bologna, Italy\\
$^{\rm IV}$ Also at: Instituto de Fisica da Universidade de Sao Paulo\\
$^{\rm V}$ Also at: Dipartimento DET del Politecnico di Torino, Turin, Italy\\
$^{\rm VI}$ Also at: Department of Applied Physics, Aligarh Muslim University, Aligarh, India\\
$^{\rm VII}$ Also at: Institute of Theoretical Physics, University of Wroclaw, Poland\\
$^{\rm VIII}$ Also at: Facultad de Ciencias, Universidad Nacional Aut\'{o}noma de M\'{e}xico, Mexico City, Mexico\\

\section*{Collaboration Institutes}

$^{1}$ A.I. Alikhanyan National Science Laboratory (Yerevan Physics Institute) Foundation, Yerevan, Armenia\\
$^{2}$ AGH University of Krakow, Cracow, Poland\\
$^{3}$ Bogolyubov Institute for Theoretical Physics, National Academy of Sciences of Ukraine, Kiev, Ukraine\\
$^{4}$ Bose Institute, Department of Physics  and Centre for Astroparticle Physics and Space Science (CAPSS), Kolkata, India\\
$^{5}$ California Polytechnic State University, San Luis Obispo, California, United States\\
$^{6}$ Central China Normal University, Wuhan, China\\
$^{7}$ Centro de Aplicaciones Tecnol\'{o}gicas y Desarrollo Nuclear (CEADEN), Havana, Cuba\\
$^{8}$ Centro de Investigaci\'{o}n y de Estudios Avanzados (CINVESTAV), Mexico City and M\'{e}rida, Mexico\\
$^{9}$ Chicago State University, Chicago, Illinois, United States\\
$^{10}$ China Nuclear Data Center, China Institute of Atomic Energy, Beijing, China\\
$^{11}$ China University of Geosciences, Wuhan, China\\
$^{12}$ Chungbuk National University, Cheongju, Republic of Korea\\
$^{13}$ Comenius University Bratislava, Faculty of Mathematics, Physics and Informatics, Bratislava, Slovak Republic\\
$^{14}$ Creighton University, Omaha, Nebraska, United States\\
$^{15}$ Department of Physics, Aligarh Muslim University, Aligarh, India\\
$^{16}$ Department of Physics, Pusan National University, Pusan, Republic of Korea\\
$^{17}$ Department of Physics, Sejong University, Seoul, Republic of Korea\\
$^{18}$ Department of Physics, University of California, Berkeley, California, United States\\
$^{19}$ Department of Physics, University of Oslo, Oslo, Norway\\
$^{20}$ Department of Physics and Technology, University of Bergen, Bergen, Norway\\
$^{21}$ Dipartimento di Fisica, Universit\`{a} di Pavia, Pavia, Italy\\
$^{22}$ Dipartimento di Fisica dell'Universit\`{a} and Sezione INFN, Cagliari, Italy\\
$^{23}$ Dipartimento di Fisica dell'Universit\`{a} and Sezione INFN, Trieste, Italy\\
$^{24}$ Dipartimento di Fisica dell'Universit\`{a} and Sezione INFN, Turin, Italy\\
$^{25}$ Dipartimento di Fisica e Astronomia dell'Universit\`{a} and Sezione INFN, Bologna, Italy\\
$^{26}$ Dipartimento di Fisica e Astronomia dell'Universit\`{a} and Sezione INFN, Catania, Italy\\
$^{27}$ Dipartimento di Fisica e Astronomia dell'Universit\`{a} and Sezione INFN, Padova, Italy\\
$^{28}$ Dipartimento di Fisica `E.R.~Caianiello' dell'Universit\`{a} and Gruppo Collegato INFN, Salerno, Italy\\
$^{29}$ Dipartimento DISAT del Politecnico and Sezione INFN, Turin, Italy\\
$^{30}$ Dipartimento di Scienze MIFT, Universit\`{a} di Messina, Messina, Italy\\
$^{31}$ Dipartimento Interateneo di Fisica `M.~Merlin' and Sezione INFN, Bari, Italy\\
$^{32}$ European Organization for Nuclear Research (CERN), Geneva, Switzerland\\
$^{33}$ Faculty of Electrical Engineering, Mechanical Engineering and Naval Architecture, University of Split, Split, Croatia\\
$^{34}$ Faculty of Nuclear Sciences and Physical Engineering, Czech Technical University in Prague, Prague, Czech Republic\\
$^{35}$ Faculty of Physics, Sofia University, Sofia, Bulgaria\\
$^{36}$ Faculty of Science, P.J.~\v{S}af\'{a}rik University, Ko\v{s}ice, Slovak Republic\\
$^{37}$ Faculty of Technology, Environmental and Social Sciences, Bergen, Norway\\
$^{38}$ Frankfurt Institute for Advanced Studies, Johann Wolfgang Goethe-Universit\"{a}t Frankfurt, Frankfurt, Germany\\
$^{39}$ Fudan University, Shanghai, China\\
$^{40}$ Gangneung-Wonju National University, Gangneung, Republic of Korea\\
$^{41}$ Gauhati University, Department of Physics, Guwahati, India\\
$^{42}$ Helmholtz-Institut f\"{u}r Strahlen- und Kernphysik, Rheinische Friedrich-Wilhelms-Universit\"{a}t Bonn, Bonn, Germany\\
$^{43}$ Helsinki Institute of Physics (HIP), Helsinki, Finland\\
$^{44}$ High Energy Physics Group,  Universidad Aut\'{o}noma de Puebla, Puebla, Mexico\\
$^{45}$ Horia Hulubei National Institute of Physics and Nuclear Engineering, Bucharest, Romania\\
$^{46}$ HUN-REN Wigner Research Centre for Physics, Budapest, Hungary\\
$^{47}$ Indian Institute of Technology Bombay (IIT), Mumbai, India\\
$^{48}$ Indian Institute of Technology Indore, Indore, India\\
$^{49}$ INFN, Laboratori Nazionali di Frascati, Frascati, Italy\\
$^{50}$ INFN, Sezione di Bari, Bari, Italy\\
$^{51}$ INFN, Sezione di Bologna, Bologna, Italy\\
$^{52}$ INFN, Sezione di Cagliari, Cagliari, Italy\\
$^{53}$ INFN, Sezione di Catania, Catania, Italy\\
$^{54}$ INFN, Sezione di Padova, Padova, Italy\\
$^{55}$ INFN, Sezione di Pavia, Pavia, Italy\\
$^{56}$ INFN, Sezione di Torino, Turin, Italy\\
$^{57}$ INFN, Sezione di Trieste, Trieste, Italy\\
$^{58}$ Inha University, Incheon, Republic of Korea\\
$^{59}$ Institute for Gravitational and Subatomic Physics (GRASP), Utrecht University/Nikhef, Utrecht, Netherlands\\
$^{60}$ Institute of Experimental Physics, Slovak Academy of Sciences, Ko\v{s}ice, Slovak Republic\\
$^{61}$ Institute of Physics, Homi Bhabha National Institute, Bhubaneswar, India\\
$^{62}$ Institute of Physics of the Czech Academy of Sciences, Prague, Czech Republic\\
$^{63}$ Institute of Space Science (ISS), Bucharest, Romania\\
$^{64}$ Institut f\"{u}r Kernphysik, Johann Wolfgang Goethe-Universit\"{a}t Frankfurt, Frankfurt, Germany\\
$^{65}$ Instituto de Ciencias Nucleares, Universidad Nacional Aut\'{o}noma de M\'{e}xico, Mexico City, Mexico\\
$^{66}$ Instituto de F\'{i}sica, Universidade Federal do Rio Grande do Sul (UFRGS), Porto Alegre, Brazil\\
$^{67}$ Instituto de F\'{\i}sica, Universidad Nacional Aut\'{o}noma de M\'{e}xico, Mexico City, Mexico\\
$^{68}$ iThemba LABS, National Research Foundation, Somerset West, South Africa\\
$^{69}$ Jeonbuk National University, Jeonju, Republic of Korea\\
$^{70}$ Johann-Wolfgang-Goethe Universit\"{a}t Frankfurt Institut f\"{u}r Informatik, Fachbereich Informatik und Mathematik, Frankfurt, Germany\\
$^{71}$ Laboratoire de Physique Subatomique et de Cosmologie, Universit\'{e} Grenoble-Alpes, CNRS-IN2P3, Grenoble, France\\
$^{72}$ Lawrence Berkeley National Laboratory, Berkeley, California, United States\\
$^{73}$ Lund University Department of Physics, Division of Particle Physics, Lund, Sweden\\
$^{74}$ Nagasaki Institute of Applied Science, Nagasaki, Japan\\
$^{75}$ Nara Women{'}s University (NWU), Nara, Japan\\
$^{76}$ National and Kapodistrian University of Athens, School of Science, Department of Physics , Athens, Greece\\
$^{77}$ National Centre for Nuclear Research, Warsaw, Poland\\
$^{78}$ National Institute of Science Education and Research, Homi Bhabha National Institute, Jatni, India\\
$^{79}$ National Nuclear Research Center, Baku, Azerbaijan\\
$^{80}$ National Research and Innovation Agency - BRIN, Jakarta, Indonesia\\
$^{81}$ Niels Bohr Institute, University of Copenhagen, Copenhagen, Denmark\\
$^{82}$ Nikhef, National institute for subatomic physics, Amsterdam, Netherlands\\
$^{83}$ Nuclear Physics Group, STFC Daresbury Laboratory, Daresbury, United Kingdom\\
$^{84}$ Nuclear Physics Institute of the Czech Academy of Sciences, Husinec-\v{R}e\v{z}, Czech Republic\\
$^{85}$ Oak Ridge National Laboratory, Oak Ridge, Tennessee, United States\\
$^{86}$ Ohio State University, Columbus, Ohio, United States\\
$^{87}$ Physics department, Faculty of science, University of Zagreb, Zagreb, Croatia\\
$^{88}$ Physics Department, Panjab University, Chandigarh, India\\
$^{89}$ Physics Department, University of Jammu, Jammu, India\\
$^{90}$ Physics Program and International Institute for Sustainability with Knotted Chiral Meta Matter (WPI-SKCM$^{2}$), Hiroshima University, Hiroshima, Japan\\
$^{91}$ Physikalisches Institut, Eberhard-Karls-Universit\"{a}t T\"{u}bingen, T\"{u}bingen, Germany\\
$^{92}$ Physikalisches Institut, Ruprecht-Karls-Universit\"{a}t Heidelberg, Heidelberg, Germany\\
$^{93}$ Physik Department, Technische Universit\"{a}t M\"{u}nchen, Munich, Germany\\
$^{94}$ Politecnico di Bari and Sezione INFN, Bari, Italy\\
$^{95}$ Research Division and ExtreMe Matter Institute EMMI, GSI Helmholtzzentrum f\"ur Schwerionenforschung GmbH, Darmstadt, Germany\\
$^{96}$ Saga University, Saga, Japan\\
$^{97}$ Saha Institute of Nuclear Physics, Homi Bhabha National Institute, Kolkata, India\\
$^{98}$ School of Physics and Astronomy, University of Birmingham, Birmingham, United Kingdom\\
$^{99}$ Secci\'{o}n F\'{\i}sica, Departamento de Ciencias, Pontificia Universidad Cat\'{o}lica del Per\'{u}, Lima, Peru\\
$^{100}$ Stefan Meyer Institut f\"{u}r Subatomare Physik (SMI), Vienna, Austria\\
$^{101}$ SUBATECH, IMT Atlantique, Nantes Universit\'{e}, CNRS-IN2P3, Nantes, France\\
$^{102}$ Sungkyunkwan University, Suwon City, Republic of Korea\\
$^{103}$ Suranaree University of Technology, Nakhon Ratchasima, Thailand\\
$^{104}$ Technical University of Ko\v{s}ice, Ko\v{s}ice, Slovak Republic\\
$^{105}$ The Henryk Niewodniczanski Institute of Nuclear Physics, Polish Academy of Sciences, Cracow, Poland\\
$^{106}$ The University of Texas at Austin, Austin, Texas, United States\\
$^{107}$ Universidad Aut\'{o}noma de Sinaloa, Culiac\'{a}n, Mexico\\
$^{108}$ Universidade de S\~{a}o Paulo (USP), S\~{a}o Paulo, Brazil\\
$^{109}$ Universidade Estadual de Campinas (UNICAMP), Campinas, Brazil\\
$^{110}$ Universidade Federal do ABC, Santo Andre, Brazil\\
$^{111}$ Universitatea Nationala de Stiinta si Tehnologie Politehnica Bucuresti, Bucharest, Romania\\
$^{112}$ University of Derby, Derby, United Kingdom\\
$^{113}$ University of Houston, Houston, Texas, United States\\
$^{114}$ University of Jyv\"{a}skyl\"{a}, Jyv\"{a}skyl\"{a}, Finland\\
$^{115}$ University of Kansas, Lawrence, Kansas, United States\\
$^{116}$ University of Liverpool, Liverpool, United Kingdom\\
$^{117}$ University of Science and Technology of China, Hefei, China\\
$^{118}$ University of South-Eastern Norway, Kongsberg, Norway\\
$^{119}$ University of Tennessee, Knoxville, Tennessee, United States\\
$^{120}$ University of the Witwatersrand, Johannesburg, South Africa\\
$^{121}$ University of Tokyo, Tokyo, Japan\\
$^{122}$ University of Tsukuba, Tsukuba, Japan\\
$^{123}$ Universit\"{a}t M\"{u}nster, Institut f\"{u}r Kernphysik, M\"{u}nster, Germany\\
$^{124}$ Universit\'{e} Clermont Auvergne, CNRS/IN2P3, LPC, Clermont-Ferrand, France\\
$^{125}$ Universit\'{e} de Lyon, CNRS/IN2P3, Institut de Physique des 2 Infinis de Lyon, Lyon, France\\
$^{126}$ Universit\'{e} de Strasbourg, CNRS, IPHC UMR 7178, F-67000 Strasbourg, France, Strasbourg, France\\
$^{127}$ Universit\'{e} Paris-Saclay, Centre d'Etudes de Saclay (CEA), IRFU, D\'{e}partment de Physique Nucl\'{e}aire (DPhN), Saclay, France\\
$^{128}$ Universit\'{e}  Paris-Saclay, CNRS/IN2P3, IJCLab, Orsay, France\\
$^{129}$ Universit\`{a} degli Studi di Foggia, Foggia, Italy\\
$^{130}$ Universit\`{a} del Piemonte Orientale, Vercelli, Italy\\
$^{131}$ Universit\`{a} di Brescia, Brescia, Italy\\
$^{132}$ Variable Energy Cyclotron Centre, Homi Bhabha National Institute, Kolkata, India\\
$^{133}$ Warsaw University of Technology, Warsaw, Poland\\
$^{134}$ Wayne State University, Detroit, Michigan, United States\\
$^{135}$ Yale University, New Haven, Connecticut, United States\\
$^{136}$ Yildiz Technical University, Istanbul, Turkey\\
$^{137}$ Yonsei University, Seoul, Republic of Korea\\
$^{138}$ Affiliated with an institute formerly covered by a cooperation agreement with CERN\\
$^{139}$ Affiliated with an international laboratory covered by a cooperation agreement with CERN.\\

\end{flushleft}

\end{document}